\documentclass[11pt, onecolumn, draftcls]{IEEEtran}

\usepackage{graphicx} 
\usepackage{amssymb}
\usepackage{verbatim}
\usepackage{graphicx}
\usepackage{amsmath}
\usepackage{url}
\usepackage{amsthm, bm}
\usepackage{enumerate}
\usepackage[square,sort&compress,comma,numbers]{natbib}
\usepackage{cleveref}
\usepackage{float}

\newcommand{\Lc}{\mathcal{L}}
\newcommand{\Ed}{{{\mathbb E}}}
\usepackage{dblfloatfix}
\usepackage[font=small,labelfont=bf]{caption}
\usepackage{color}
\usepackage{booktabs}
\usepackage{wrapfig}

\usepackage{xcolor}
\usepackage[]{color-edits}
\usepackage[normalem]{ulem}
\newcommand{\changedel}[1]{{\color{blue}{\sout{#1}}}}

\usepackage{xargs}  
\usepackage[colorinlistoftodos,prependcaption,textsize=tiny]{todonotes}
\makeatletter
\def\th@plain{%
  \thm@notefont{}% same as heading font
  \itshape % body font
}
\def\th@definition{%
  \thm@notefont{}% same as heading font
  \normalfont % body font
}
\makeatother

\newcommand{\nfin}{N_{f,\text{in}}}
\newcommand{\nfout}{N_{f,\text{out}}}
\newcommand{\x}{\mathbf{x}}
\newcommand{\Real}[1]{Re(#1)}
\newcommand{\Imag}[1]{Im(#1)}
\newcommand{\abs}[1]{\lvert #1 \rvert}
\newcommand{\fracdd}[2]{\frac{\partial #1}{\partial #2}}

\usepackage{xargs}

\begin{document}
\title{Physics-Driven Deep Learning for Computational Magnetic Resonance Imaging}
%\markboth{WHITE PAPER}{WHITE PAPER}

\author{Kerstin Hammernik, Thomas K\"ustner, Burhaneddin Yaman, Zhengnan Huang, \\
Daniel Rueckert, Florian Knoll, Mehmet Ak\c{c}akaya
	\vspace{-0.3cm}
	\thanks{K. Hammernik and D. Rueckert are with the Institute of AI and Informatics in Medicine, Technical University of Munich and the Department of Computing, Imperial College London.}
	\thanks{T. K\"ustner is with the Department of Diagnostic and Interventional Radiology, University Hospital of Tuebingen.}
	\thanks{B. Yaman and M. Ak\c{c}akaya are with the Department of Electrical and Computer Engineering, and Center for Magnetic Resonance Research, University of Minnesota, USA.}
	\thanks{Z. Huang is with the Center for Biomedical Imaging, Department of Radiology, New York University.}
	\thanks{F. Knoll is with the Department Artificial Intelligence in Biomedical Engineering,
           Friedrich-Alexander University Erlangen.}
	\thanks{We acknowledge grant support from the NIH, Grant Numbers: R01HL153146, P41EB027061, R01EB024532 and P41EB017183; from the NSF, Grant Number: CAREER CCF-1651825; and from the EPSRC Programme Grant, Grant Number: EP/P001009/1.}}
	
\maketitle

\begin{abstract}
Physics-driven deep learning methods have emerged as a powerful tool for computational magnetic resonance imaging (MRI) problems, pushing reconstruction performance to new limits. This article provides an overview of the recent developments in incorporating physics information into learning-based MRI reconstruction. We consider inverse problems with both linear and non-linear forward models for computational MRI, and review the classical approaches for solving these. We then focus on physics-driven deep learning approaches, covering physics-driven loss functions, plug-and-play methods, generative models, and unrolled networks. We highlight domain-specific challenges such as real- and complex-valued building blocks of neural networks, and translational applications in MRI with linear and non-linear forward models. Finally, we discuss common issues and open challenges, and draw connections to the importance of physics-driven learning when combined with other downstream tasks in the medical imaging pipeline.
\end{abstract}

\IEEEpeerreviewmaketitle	

%%%%%%%%%%%%%%%%%%%%%%%%%%%%%%%%%%%%%%%%%%%%%%%%%%%%%%%%%%%%%%%%%%%%%%%%%%%%%%%%
\section{Introduction}
\label{sec:Introduction}
Magnetic resonance imaging (MRI) is a non-invasive radiation-free imaging modality with a plethora of clinical applications and extensively-studied physics underpinnings. {The relationship between the acquired MRI data and the underlying magnetization is characterized by Bloch equations, and depends on a number of parameters, including the magnetic fields (e.g. the static B$_0$ magnetic field), relaxation effects (e.g. T$_1$, T$_2$ relaxation), motions at different scales (e.g. physiological, flow, diffusion and perfusion), and acquisition parameters (e.g. echo time, flip angle) \cite{Doneva2020}. These intricate dependencies are encoded in the so-called k-space}, corresponding to the spatial Fourier transform of the object's magnetization. 

{The acquired k-space signal $y(t)$ at time $t$, prior to discretization, is given as
\begin{align}
    y(t) = \int M(\rho({\bf r}), {\bm \vartheta}, t, {\bf r}) \mathrm e^{-j 2\pi {\bf k}(t) \cdot {\bf r}}\, \mathrm d {\bf r} + n(t)
    \label{eq:nonlinearModelKspace}
\end{align}
where ${\bf r}$ is the spatial location; $\rho({\bf r})$ is the underlying spin densities/transverse magnetization; ${\bm \vartheta}$ is a set of (potentially unknown) parameters that model physiological or systemic changes, and themselves may depend on ${\bf r}$; $k(t)$ is the k-space location at time $t$ sampled along a k-space trajectory under the influence of spatially and temporally varying magnetic fields; and $n(t)$ is measurement noise. The physics-based signal model $M(\rho({\bf r}), {\bm \vartheta}, t_j, {\bf r})$, sampled at times $t_j$, thereby describes the effects that influence the underlying magnetization, based on pre-specified and known image acquisition parameters. It depends on the imaging sequence and reflects physiological, functional or hardware characteristics. For many applications, an analytical expression can be derived (e.g. via hard pulse approximation from the Bloch equations) for which a few examples are summarized in Table \ref{tab:PhysicsbasedModels} (linear and non-linear models). If no analytical expression can be derived for the imaging sequence, the Bloch equations need to be integrated directly as the signal model.

\begin{table}[h]
    \centering
    \caption{Analytically derived physics-based signal models for pre-specified imaging sequences for set and known image acquisition parameters (dominant one influencing the signal model is depicted) and the to be estimated unknown parameters \cite{Doneva2020, Odille2008,  seiberlich2020quantitative, batchelor2005matrix}.}
    \begin{tabular}{lccc}
        \toprule
        physical effect & image acquisition parameters & unknown parameter ${\bm \vartheta}$ & signal model $M(\rho({\bf r}), {\bm \vartheta}, t_j, {\bf r})$ \\ \midrule
        off-resonance & echo time $t_j$ & $\Delta\omega$ & $\mathrm e^{j \Delta \omega({\bf r}) t_j} \rho({\bf r})$ \\
        motion & echo time $t_j$ & motion field ${\bf U}_{j}$ & $\rho\left({\bf U}_{j}(\bf r)\right) |\det(\nabla {\bf U}_{j})(\bf r)|$ \\
        T$_1$ relaxation & \shortstack{inversion times $t_k$, equilibrium magnetization $\rho_0$} & $T_1$ & $\rho_{0}({\bf r}) (1- \mathrm e^{-\frac{t_j}{T_1(\bf r)}})$ \\
        T$_2$ relaxation & echo time $t_j$ & $T_2$ & $\mathrm e^{-\frac{t_j}{T_2(\bf r)}} \rho({\bf r})$ \\
        flow velocity ${\bf v}$ & flow-encoded acquisitions ${\bf V}_j$ & ${\bf v}$ & $\mathrm e^{j {\bf v} \cdot {\bf V}_j} \rho({\bf r})$ \\
        diffusion tensor ${\bf D}$ & diffusion-encoded acquisitions ${\bf b}_j$ & ${\bf D}$ & $\mathrm e^{-{\bf b}_j^T {\bf D} {\bf b}_j} \rho({\bf r})$ \\ \bottomrule
    \end{tabular}
   
    \label{tab:PhysicsbasedModels}
\end{table}

For a simplified acquisition model, the signal in $M(\rho({\bf r}), {\bm \vartheta}, t, {\bf r})$ is often characterized as $x({\bf r})$, which absorbs the dependencies on the physiologic or systemic effects, as well as the signal evolution (or time-course), into the image/magnetization of interest. For example, this type of simplification is used when referring to images with different contrast weightings, such as T$_1$ or T$_2$ weighting. In this setup, following discretization, the physics-based forward model becomes linear and can be expressed as}
\begin{equation} \label{eq:forwardbasic}
    {\bf y} = {\bf E} \x + {\bf n},
\end{equation}
where ${\bf x} \in \mathbb{C}^{n}$ is {the image/magnetization of interest, ${\bf y}{\in \mathbb{C}^{m}}$ denotes the corresponding k-space measurements}, ${\bf E}{:\mathbb{C}^{n} \to \mathbb{C}^{m}}$ is the forward MRI encoding operator, and ${\bf n}{\in \mathbb{C}^{m}}$ is {discretized} measurement noise. 

In its simplest form, ${\bf E}$ corresponds to a sub-sampled discrete Fourier transform matrix ${\bf F}_{\Omega}{: \mathbb{C}^{n}\to \mathbb{C}^{m}}$ which samples the k-space locations specified by $\Omega$. In practice, however, all clinical MRI scanners from all vendors are equipped with multi-coil receiver arrays, and the corresponding multi-coil forward operator ${\bf E}{:\mathbb{C}^{n}\to\mathbb{C}^{m\cdot n_c}}$ is given as 
$${\bf E} = \left[\begin{array}{c} {\bf F}_{\Omega}{\bf C}_1  \\
\vdots \\
{\bf F}_{\Omega}{\bf C}_{n_c}  
\end{array}\right],$$
where $n_c$ is the number of coils in the receiver array, and ${\bf C}_{{q}}{:\mathbb{C}^{n}\to\mathbb{C}^{n}}$ is a diagonal
matrix containing the sensitivity profile of the {$q^\textrm{th}$} receiver coil. {These coil sensitivities are typically pre-estimated from subject-specific calibration data \cite{Pruessmann1999}.} {We note that while ${\bf F}_{\Omega}$ typically refers to a sub-sampled Cartesian acquisition that can be implemented efficiently with a fast Fourier transform, non-Cartesian acquisitions are also used in some clinical applications \cite{Seiberlich2014}.}

Formation of images and other information from these measured k-space data constitutes the basis of computational MRI, which in itself has a rich history \cite{Liang1992}. The canonical inverse problem for computational MRI relates to the formation of images from sub-sampled/degraded k-space data. Solving such inverse problems often necessitates incorporation of additional information about MRI encoding and/or the nature of MR images. Earlier works concentrated on the properties of the k-space, such as partial Fourier imaging methods that utilize Hermitian symmetry. With the advent of multi-coil receiver arrays, the redundancies among these coil elements became the important information for the next generation of inverse problems \cite{Pruessmann1999}. 
Subsequently, compressed sensing methods \cite{Lustig2007} were proposed to utilize the compressibility of MR images, often in addition to the redundancies among the coil elements.

In addition to the above canonical linear inverse problems, there is a class of computational MRI methods that deal{s} with {the} more complicated non-linear forward models incorporating physical, systemic and physiological parameters, {as stated in Table \ref{tab:PhysicsbasedModels}}. {The forward model in this case can be broadly given as \cite{Doneva2020}: 
\begin{equation} \label{eq:forwardnonlinear}
  {  {\bf y} = {\cal E}({\bf v}) + {\bf n},}
\end{equation}
where ${\bf v} \in {\mathbb C}^{n_v}$ is a vector that includes all unknown imaging/quantity  and parameters of interest that describes the signal evolution in Eq. \eqref{eq:nonlinearModelKspace}, and ${\cal E}: \mathbb{C}^{n_v} \to \mathbb{C}^{m\cdot n_c}$ is a non-linear encoding operator, i.e. the signal evolution arising from the physics-based signal model of Eq. \eqref{eq:nonlinearModelKspace}. {It can be decomposed into ${\cal E} = {\bf E}{\cal M}$, where ${\bf E}$ is the canonical multi-coil forward operator and ${\cal M}: \mathbb{C}^{n_v n_\vartheta} \to \mathbb{C}^n$ is the discretized signal model describing the spin physics.} Here, we make the distinction that ${\bf v}$ includes all unknown quantities of interest that describes the signal evolution, as opposed to just an image as in Eq. \eqref{eq:forwardbasic}. This broad definition is necessary to incorporate different setups \cite{Doneva2020}, which are partially described in Table \ref{tab:PhysicsbasedModels}. For example, for the motion model in Table \ref{tab:PhysicsbasedModels}, ${\bf v}$ includes both the motion field and the image of interest. For this model, the former was specified by ${\bm \vartheta}$ as the unknown physiological parameter, but one is typically interested in recovering the image itself.
For a relaxation model, ${\bf v}$ includes both the magnetization and the relaxation map (e.g. T$_1$ or T$_2$). In this setup, the quantity of interest is the relaxation map, which was specified by ${\bm \vartheta}$ as the unknown physical parameter, but the magnetization also needs to be recovered to fully describe the model. Thus, it is not straightforward to tease out the unknown parameter from the magnetization in all cases, where the object of clinical interest may be either. Hence the unified notation \cite{Doneva2020} makes the inverse problem easier to specify without going into application details. Traditionally, the inverse problem corresponding to Eq. \eqref{eq:forwardnonlinear} is solved using model-based reconstructions.}

Recently, deep learning methods have emerged as a powerful tool for solving many inverse problems in computational MRI. Among these, MRI reconstruction for accelerated acquisitions remains the most well-studied \cite{Hammernik2018,Aggarwal2019,Knoll_SPM}, along with several strategies for quantitative MRI \cite{liu2019mantis}, 
motion \cite{KuestnerAPSIPA2020,Qi2021} and other non-linear physical models \cite{yoon2018quantitative, Fathi2020}. {Out of a plethora of approaches for these problems, physics-driven methods,which explicitly use the known physics-based forward imaging models in deep learning architectures and/or training to control the consistency of the reconstruction with k-space measurements, have emerged as the most well-received deep learning techniques by the MRI community due to their incorporation of the MR domain knowledge.} The goal of this manuscript is to provide a comprehensive review of inverse problems for computational MRI, and how physics-driven deep learning techniques {involving the raw k-space data} are being used for these applications.

\section{Classical approaches for computational MRI}
The simplest image reconstruction problem for computational MRI concerns the case where ${\bf E}$ is exactly the discrete Fourier transform matrix in Eq. \eqref{eq:forwardbasic}, corresponding to Nyquist rate sampling for a given resolution and field-of-view. In this case, the image of interest is recovered via inverse discrete Fourier transform.

In practical settings, often a sub-Nyquist rate is employed to enable faster imaging, where the previous simple strategy of taking the inverse Fourier transform leads to aliasing artifacts. Thus, in this regime, an inverse problem, incorporating additional domain knowledge, needs to be solved for image formation. The most commonly used clinical strategy for accelerated MRI is parallel imaging \cite{Pruessmann1999}, which uses the redundancies among these coil elements for image reconstruction. Succinctly, parallel imaging methods that work in image domain \cite{Pruessmann1999} solved 
\begin{equation} \label{eq:LS}
    \hat{\bf x}_{PI} =  \arg \min_{\bf x} \frac12 ||{\bf y - Ex}||_2^2 = ({\bf E}^H{\bf E})^{-1}{\bf E}^H{\bf y}{,}
\end{equation}
{where $H$ denotes the Hermitian transpose.}
In theory, with $n_c$ coil elements, the ratio between the image size and the cardinality of $\Omega$, or the acceleration rate ($R$), can be as high as $n_c$. However, due to spatial or statistical dependencies between $\{{\bf C}_k\}$ and ill-conditioning of ${\bf E}$ that leads to noise amplification due to the matrix inversion \cite{Pruessmann1999}, the achievable rates are often limited. Subsequently, compressed sensing methods \cite{Lustig2007} were proposed to utilize the compressibility of MR images to reconstruct images from sub-sampled k-space data. These methods solve a regularized least squares objective function
\begin{equation} \label{eq:CS}
    \hat{\bf x}_{CS} = \arg \min_{\bf x} \frac12 ||{\bf y - Ex}||_2^2 + \tau ||{\bf Wx}||_1,
\end{equation}
where $||\cdot||_1$ denotes the $\ell_1$ norm, ${\bf W}$ is a sparsifying linear transform, and $\tau$ is a weight term. Unlike (\ref{eq:LS}), the objective function does not have a closed form solution. {We also note that Eq. \eqref{eq:CS} corresponds to the analysis formulation of $\ell_1$ regularization \cite{Candes2011}, while the synthesis formulation which performs regularization in the transform domain directly also remains popular. The two are equivalent when ${\bf W}$ is a unitary transformation. Both the synthesis and analysis formulations lead to a convex problem, and can be solved using numerous iterative algorithms \cite{Fessler2020}. }

\subsection{Solving the linear inverse problem in classical computational MRI} \label{sec:2a}
In general, we will consider a regularized least squares objective with a broader class of regularizers:
\begin{equation}
    \hat{\bf x}_{reg} = \arg \min_{\bf x} \frac12 ||{\bf y - Ex}||_2^2 + {\cal R}({\bf x}),\label{eq:regularized_problem}
\end{equation}
where ${\cal R}(\cdot)$ may be one of the aforementioned regularizers, such as the $\ell_1$ norm of transform domain {coefficients \cite{UnserBook}}, or implicitly implemented via machine learning techniques, as we will later see.

There is a number of iterative algorithms for solving such objective functions, especially when it is convex \cite{Fessler2020}. A classical approach, when ${\cal R}(\cdot)$ is differentiable, is based on gradient descent:
\begin{equation} \label{eq:gd-iters}
    {\bf x}^{(i)} ={\bf x}^{(i-1)}+\eta {\bf E}^H({\bf y - Ex}^{(i-1)}) - \eta \nabla_{\bf x} {\cal R}({\bf x})|_{{\bf x = x}^{(i-1)}},
\end{equation}
where ${\bf x}^{(i)}$ is the image of interest at the $i^\textrm{th}$ iteration. However, often times nonsmooth regularizers are used in computational MRI. In this case, proximal algorithms are widely used \cite{Fessler2020}. One such method is proximal gradient descent, which amounts to solving two sub-problems:
{{
\begin{subequations}
\begin{align}
{\bf z}^{(i)} &= {\bf x}^{(i-1)}+\eta {\bf E}^H({\bf y - E}{\bf x}^{(i-1)}), \label{eq:pgd-updates-dc}\\
{\bf x}^{(i)} &= \arg \min_{\bf x} \frac12 || {\bf z}^{(i)} - {\bf x}||_2^2 + \eta {\cal R}({\bf x}) \triangleq \textrm{prox}_{{\cal R},\eta}({\bf z}^{(i)}) \label{eq:pgd-updates-prox}
\end{align}
\end{subequations}}}
where ${\bf x}^{(i)}$ and ${\bf z}^{(i)}$ are the image of interest and an intermediate image at the $i^\textrm{th}$ iteration respectively, Eq. (\ref{eq:pgd-updates-prox}) corresponds to the so-called proximal operator for the regularizer, Eq. (\ref{eq:pgd-updates-dc}) encourages data consistency, and $\eta$ is a step size. 

Another class of popular approaches rely on variable splitting, such as the alternating direction method of multipliers (ADMM), which solves:
\begin{subequations}
\begin{align}
{\bf x}^{(i)} &= \big({\bf E}^H{\bf E} + \rho {\bf I}\big)^{-1} \big({\bf E}^H{\bf y} + \rho({\bf z}^{(i-1)} - {\bf u}^{(i-1)})\big),  \label{eq:admm-dc}\\
{\bf z}^{(i)}&=\arg \min_{\bf z} \frac12 \big|\big|({\bf x}^{(i)} + {\bf u}^{(i-1)}) - {\bf z}\big|\big|_2^2 + \frac{1}{\rho} {\cal R}({\bf z}), \label{eq:admm-prox} \\
{\bf u}^{(i)} &= {\bf u}^{(i-1)} + ({\bf x}^{(i)} - {\bf z}^{(i)} ), \label{eq:admm-dual}
\end{align}
\end{subequations}
where ${\bf x}^{(i)}$ is the image of interest at the $i^\textrm{th}$ iteration, ${\bf z}^{(i)}$ and ${\bf u}^{(i)}$ are intermediate images, and $\rho$ is a penalty weight. Here, (\ref{eq:admm-dc}), (\ref{eq:admm-prox}) and (\ref{eq:admm-dual}) corresponds to data consistency, proximal operator and dual update sub-problems respectively. A simpler version of variable splitting is based on a quadratic penalty \cite{Fessler2020}, which leads to the following equations:
{
\begin{subequations}
\begin{align}
{\bf x}^{(i)} &= \big({\bf E}^H{\bf E} + \rho {\bf I}\big)^{-1} \big({\bf E}^H{\bf y} + \rho {\bf z}^{(i-1)} \big),  \label{eq:vsqp-dc}\\
{\bf z}^{(i)}&=\arg \min_{\bf z} \frac12 \big|\big|{\bf x}^{(i)}  - {\bf z}\big|\big|_2^2 + \frac{1}{\rho} {\cal R}({\bf z}), \label{eq:vsqp-prox} 
\end{align}
\end{subequations}
}
\subsection{Solving the non-linear inverse problem in classical computational MRI}

{The general unconstrained optimization problem for a model-based reconstruction for the forward model in Eq. \eqref{eq:forwardnonlinear} 
can be stated as:
\begin{align}
    \hat{\bf v} = \arg\min_{{\bf v}} \|{\bf E}{\cal M}({\bf v}) - {\bf y}\|_2^2 + R_{\bf v}({\bf v})  \label{eq:nonlinearSol} 
\end{align}
where $R_{\bf v}$ is a (combination of) regularizer that acts on all unknown quantities of interest. While the notation is general, the regularizer may be separable among different quantities, e. g. different regularizers for the motion field and the image of interest.}

This description can be used to combine parallel imaging, compressed sensing and model-based reconstruction in a unified formulation.  
{In addition to the motion and T$_2$ mapping models discussed earlier in Section \ref{sec:Introduction}, a non-linear forward model can also be used to describe dynamic imaging scenarios, such as contrast-enhanced imaging. Consider the signal model in Eq. \eqref{eq:nonlinearModelKspace}, and time instances specified $\tau_0(t) \triangleq 0, \tau_1(t), \tau_2(t), \dots, \tau_n(t)$. These time instances may correspond to different physical events, e.g. RF excitation for single-shot EPI acquisitions, sampling after an inversion pulse for T$_1$ mapping, or cardiac triggering for myocardial parameter mapping or perfusion cardiac MRI. Let the discretized k-space measurements between $\tau_{i-1}(t)$ and $\tau_{i}(t)$ be denoted by ${\bf y}_i$. Thus, each ${\bf y}_i$, corresponding to $\{y(t)\}_{\tau_{i-1}(t) \leq t < \tau_i(t)}$, essentially captures a snap-shot of this dynamic process between the specified sample instances. In the same vein as Eq. \eqref{eq:forwardnonlinear}, these can be vectorized into ${\bf y}$, where the corresponding ${\bf v}$ models the relevant pharmacokinetic quantities.}

For the inverse problems with non-linear forward operators, the algorithms are less standardized, and typically application-dependent. Eq. \eqref{eq:nonlinearSol} is usually non-convex, making its optimization a challenging task. Furthermore, inaccuracies or incompleteness of the modelling can further influence the optimization. One approach is to employ a Gauss-Newton algorithm, and linearize the problem around the solution of the previous iteration \cite{Liang1992} or by approximating the non-linear behaviour with a linear combination of basis functions \cite{Asslander2018}.

\section{Physics-driven ML methods in computational MRI}
Deep learning methods have recently emerged as a powerful tool for computational MRI. These methods can be broadly split into two classes: purely data-driven and physics-driven~\cite{Liang2020}. The former methods are typically implemented in image space, as removing artifacts from aliased images \cite{Wang2016}. These image enhancement networks are typically trained to map corrupted and undersampled images to artifact-free images. Indeed, learning image enhancement networks is the key ingredient to remove artifacts in image domain. However, when only image enhancement methods are used, the information of the acquisition physics is entirely discarded, hence, k-space consistency cannot be guaranteed. In this section, we will give an extensive overview on physics-driven deep learning methods for computational MRI, ranging from physics-informed enhancement methods to learned unrolled optimization, as well as reconstruction with generative models and plug-and-play priors. 

\subsection{Physics Information in Image/k-space Enhancement Methods} \label{sec:3a}
As aforementioned, image enhancement networks typically learn a mapping from the aliased/degraded image, such as the zero-filled reconstruction, to a reference image, without consideration of the measured k-space data during the reconstruction process. {For Cartesian sampling,} {s}everal attempts have been made to incorporate physics information in this line of work \cite{Hyun2018,Yang2017} {, including enforcing k-space consistency directly after image enhancement \cite{Hyun2018}, or adding k-space consistency as an additional cost function term during training \cite{Yang2017}. The former approach directly replaces the measured k-space lines, which may lead to artifacts, while the latter cannot guarantee k-space consistency during inference, especially for cases with unseen pathologies.}
{Specifically, in \cite{Hyun2018}, enforcing hard k-space consistency directly after image enhancement was proposed, where the enhanced image was transformed into Fourier space, and the points at the sampled locations were replaced by the original k-space measurements. {However, we note that this approach cannot be applied to more complex sampling trajectories in non-Cartesian imaging~\cite{Seiberlich2014}.} {A similar approach was also used in a cross-domain approach that included both k-space and image enhancement networks \cite{Eo2018}, interleaved by an analogous data consistency operation.} In \cite{Yang2017}, k-space consistency was added as an additional cost function term during training. However, this approach cannot guarantee k-space consistency during inference, especially for cases with unseen pathologies.}
Similarly, enhancement has been proposed in k-space, as a method of interpolation \cite{Akcakaya2019}, where a non-linear interpolation function is learned from calibration data. This can be seen as an extension to the linear convolution kernels used in {generalized autocalibrating partially parallel acquisitions (GRAPPA)~\cite{Griswold2002}}. As only the calibration data is required for training, this approach can be used when large training databases are not available, but its performance may be limited at high acceleration rates where the calibration data may be insufficient \cite{Knoll_SPM}. {An alternative k-space based strategy was proposed in \cite{grappanet}, which first uses scan-specific GRAPPA kernels, followed by image refinement in both image and k-space domains.}

\subsection{Plug-and-play Methods with Deep Denoisers}
Plug-and-play (PnP) algorithms decouple image modeling from the physics of the MRI acquisition, by noting that the proximal operators in Eq. (\ref{eq:pgd-updates-prox}) or Eq. (\ref{eq:admm-prox}) correspond to conventional denoising problems \cite{Ahmad2020}. In the proximal-based formulation, these proximal denoisers are replaced by other powerful denoising algorithms, which do not necessarily have a corresponding closed form ${\cal R}(\cdot)$ expression, such as BM3D \cite{Ahmad2020}. A related approach is the regularization by denoising (RED) framework, which considers finding ${\bf x}$ that solves the {first-order} optimality condition
\begin{equation}
    {\bf 0} = {\bf E}^H({\bf Ex-y}) + {\lambda} ({\bf x} - d({\bf x})),
\end{equation}
where $d(\cdot)$ is the plug-in denoiser \cite{Ahmad2020}{, and $\lambda > 0$ denotes the regularization parameter}. The advantage of the RED formulation is that under certain conditions, the regularizer, ${\cal R}(\cdot)$ can be explicitly tied to the denoiser, $d(\cdot)$. {We refer the reader to a comprehensive review article on the subject \cite{Ahmad2020} for more details.} {We also note that, beyond the computational MRI community, there has been work characterizing the guaranteed convergence of plug-and-play networks \cite{bohra2021learning}}.

\begin{figure}[htp]
\includegraphics[width=\textwidth, height=\textheight, keepaspectratio]{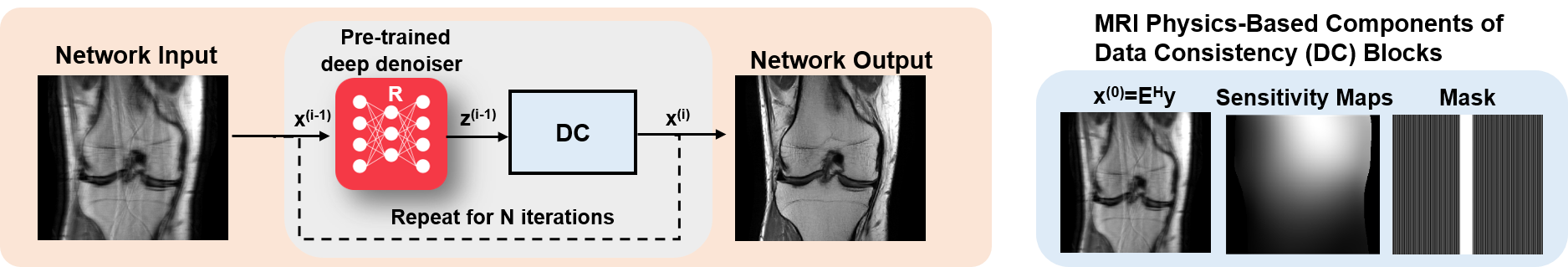}
\caption{Overview of the PnP framework in physics-driven deep learning methods for computational MRI. {The data consistency layer enforces fidelity with k-space measurements based on the known forward model. Note that a Cartesian sampling scheme is shown for easier depiction, but data consistency is also applicable to non-Cartesian trajectories.}}\label{fig:PnP}
\end{figure}

Recently, more effort has been made towards implementing CNN-based denoisers in these PnP frameworks \cite{Ahmad2020,rare, qmodel}, depicted in Figure \ref{fig:PnP}. These denoisers are typically trained using reference images in a supervised manner, where different levels of noise are retrospectively added to these images, and a mapping from the noisy images to reference images are learned \cite{Ahmad2020}. In applications, where reference images are unavailable, denoising frameworks have been proposed for training using pairs of noisy images \cite{N2N}. Extending on these works, regularization by artifact removal (RARE) trained CNN denoisers on a database of pairs of images with artifacts generated from non-Cartesian acquisitions \cite{rare}. These pairs were generated by splitting the acquired measurements in half, and reconstructing these with least squares as in Eq.\eqref{eq:LS}, corresponding to a parallel imaging reconstruction, which led to starting images of sufficient quality for non-Cartesian trajectories that oversample the central {region of} k-space. The appeal of these methods is that the CNN-based denoisers are trained independently of the broader inverse problem. Thus, only the denoising network has to be stored in memory, allowing for easier translation to larger-scale {as proposed in RARE \cite{rare} for} 3D MRI datasets . This approach is also appealing since only one denoiser has to be trained on any data. Hence,  this denoiser can, in principle, be applied across different rates or undersampling patterns. In practice, it is beneficial to provide the denoiser with additional information, such as the undersampling artifacts arising from uniform undersampling pattern in order to recognize characteristic aliasing artifacts. 

\subsection{Generative Models}
While we have reviewed explicit regularization in Eq.~\eqref{eq:regularized_problem}, regularization can also be achieved by an implicit prior in order to constrain the solution space for our optimization problem. This concept is proposed by Deep Image Prior (DIP)~\cite{Ulyanov2020} as follows:
\begin{align}
\min_{\bm \theta} \frac12 \Vert {\bf E}G_{\bm \theta}({\bf z}) - {\bf y} \Vert_2^2, \label{eq:dip}
\end{align}
where a generator network, $G_{\bm \theta}{:\mathbb{C}^{d} \to \mathbb{C}^{n}}$ parametrized by ${\bm \theta} {\in\mathbb{R}^{p}}$, reconstructs an image {${\bf x}=G_{\bm \theta}({\bf z})\in\mathbb{C}^{n}$} from a random {$d$-dimensional} latent vector ${\bf z} {\in\mathbb{C}^{d}}$. This loss function is used to train the generator network with parameters $\bm \theta$. This formulation has the advantage that it works for limited (even single) datasets without ground-truth. However, early stopping has to be performed to not overfit to the noisy measurements. While DIP methods were originally proposed for general inverse problems such as denoising, superresolution and deblurring, its application to MRI reconstruction has also been studied \cite{Yazdanpanah2019}, where the zero-filled reconstruction was used instead of the latent code as an input to the generator UNet. An extension of the DIP framework to dynamic non-{C}artesian MRI was proposed in \cite{TTDIP2021}. {A mapping network first generated an expressive latent space from a fixed low-dimensional manifold, e.g. a straight-line manifold,} using fully connected layers and non-linearities. A subsequent generative CNN generates the finally reconstructed image. A wavelet-sparsity constraint on the generated image was added as regularization to Eq. \eqref{eq:dip}, to improve the efficiency of learning and avoid over-fitting to the noisy measurements, and was further conditioned with a high-resolution image of similar anatomy~\cite{Zhao2021}. 

An alternative line of work is based on generative adverserial networks (GANs), where a generator and a discriminator network play a minimax game. The generator network samples from a fixed distribution in latent space such as Gaussian distribution and aims to map the sampling to a real data distribution in {the} ambient image space. Conversely, the discriminator network aims to differentiate between generated and real samples. The minimax training objective is defined as 
\begin{equation}
    \min_{{\bm \theta}_G} \max_{{\bm \theta}_D} \Lc_{GAN}({\bm \theta}_D, {\bm \theta}_G) \triangleq \Ed_{\bf x}[\log D_{{\bm \theta}_D}({\bf x})] + \Ed_{{\bf z}}[\log (1 - D_{{\bm \theta}_D}(G_{{\bm \theta}_G}( {\bf z}))],
\label{eq:minimax}
\end{equation} 
where the distribution on ${\bf x}$ is the real data distribution, whereas the one on ${\bf z}$ is a fixed distribution on the latent space{, and $\Ed_{{\bf z}}$ and $\Ed_{{\bf x}}$ denote the expected values defined over the random variables ${{\bf z}}$ and ${{\bf x}}$}. The generator $G_{{\bm \theta}_G}$, parametrized by {%a $p_G$-dimensional feature vector} 
${\bm \theta}_G{\in\mathbb{R}^{p_G}}$}, tries to map samples from the latent space to samples from the ambient image space, and the discriminator $D_{{\bm \theta}_D}$, parametrized by {%a $p_D$-dimensional feature vector} 
${\bm \theta}_D{\in\mathbb{R}^{p_D}}$}, tries to differentiate between the generated and the real samples.

\begin{figure}[htp]
\includegraphics[width=\textwidth, height=\textheight, keepaspectratio]{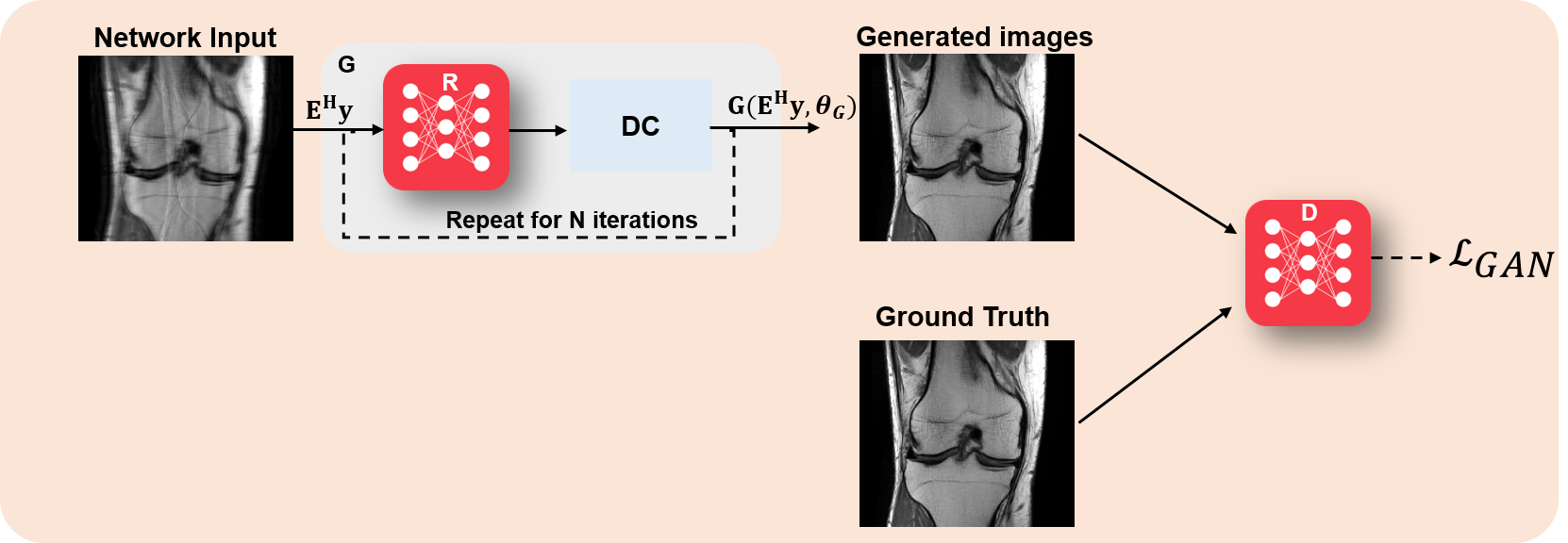}
\caption{Overview of GAN methods in physics-driven deep learning methods for computational MRI. {A generator network ($G$), typically followed by a data consistency layer, implemented using a gradient descent step as in Eq. \eqref{eq:pgd-updates-dc}, is used to generate an image. In the supervised setting, this generator is jointly trained with a discriminator network ($D$) that implements an adversarial loss to aid in the recovery of fine details of the image.} In the unsupervised setting, such as cycleGANs, physics information is further enforced in the loss function both in image and k-space domains. {Note that the figure shows the training phase, and at inference time, only the $G$ network is used.}}\label{fig:Gan}
\end{figure}

% \changeadd{The dimensions $p_G$ and $p_D$ should be selected in the same order to stabilize training.}
The idea of using GANs in computational MRI was first proposed in \cite{Mardani2018}. In this case, the generator network used the zero-filled images as input instead of a random distribution, {leading to the loss function
\begin{equation}
    \min_{{\bm \theta}_G} \max_{{\bm \theta}_D} \Ed_{\bf x}[\log D_{{\bm \theta}_D}({\bf x})] + \Ed_{{\bf {y}}}[\log (1 - D_{{\bm \theta}_D}(G_{{\bm \theta}_G}( {\bf E}^H{\bf y}))].
\label{eq:minimax-mri}
\end{equation} 
At inference time, the generator was used to produce the desired output.} Here {$G_{{\bm \theta}_G}$} was an image enhancement network, followed by a data consistency step, {for instance implemented by a gradient descent step as in Eq. \eqref{eq:pgd-updates-dc}}{; while the discriminator $D_{{\bm \theta}_D}$ was essentially used to implement an adversarial loss term to improve the recovery of finer details. Thus, this formulation used supervised training with paired data}. A high-level overview of this approach is shown in Figure \ref{fig:Gan}. A more recent work replaced this generator with a variational autoencoder based generator that also allowed for uncertainty quantification \cite{Edupuganti2021}.

Another approach is based on inverse GANs, which utilize generative learning, followed by optimization similar to the DIP~\cite{Narnhofer2019}. First, a GAN is trained to generate {an} MR image\changedel{s} from a latent noise vector. The GAN does not involve any physics-based knowledge, as only clean MRI reference images are used for training. The physics-based information and the trained generator network $G_{{\bm \theta}_G}$ are then included in {the optimization problem
\begin{align}
    \min_{{\bf z}} \frac12 \Vert {\bf E}G_{{\bm \theta}_G}({\bf z}) - {\bf y} \Vert_2^2. \label{eq:dip_z}
\end{align}
This solves for the latent vector ${\bf z}$ which is bounded from above by a hypersphere constraint, generating an image that lies in the range space of the generator. In a final step, both the generator parameters and the latent vector are optimized following:}
\begin{align}
    \min_{{\bm \theta}_G, {\bf z}} \frac12 \Vert {\bf E}G_{{\bm \theta}_G}({\bf z}) - {\bf y} \Vert_2^2, \label{eq:dip_ztheta}
\end{align}
This allows for adaptation of the generator to the undersampled k-space data at test time, and is not restricted to any sampling pattern.
{
Initialization of \eqref{eq:dip_ztheta} by the optimal latent vector found in \eqref{eq:dip_z} and early stopping (before reaching the minimum), allow the generator parameters ${\bm \theta}_G$ to not deviate too far from the original generator parameters. The reconstructed image ${\bf x}$ is obtained by using the optimized values ${\bm \theta}_G^*, {\bf z}^*$ for the generator, i.e., ${\bf x} = G_{{\bm \theta}_G^*}({\bf z}^*)$.}

In another line of work, cycle consistent GANs (cycleGAN) that enable unpaired image-to-image translation, have been analyzed using optimal transport, which provides a means to transport probability measures by minimizing average transport between measures \cite{cycleGan_Jong_MRI}. {While traditional GANs and DIP-like networks are trained to minimize a distance measure in either image space or k-space, CycleGAN aims to minimize this in both k-space and image domain.} In essence, this is achieved by minimizing two forms of losses, one for cyclic consistency and one for GAN training. 
{The former is given by}
\begin{equation}
    {\cal L}_\textrm{cycle}({\bm \theta}_G) =  \Ed_{\bf x}[||{\bf x} - G_{{\bm \theta}_G}({\bf Ex})||_2^2] + \Ed_{\bf y}[||{\bf y} - {\bf E} G_{{\bm \theta}_G}({\bf y})||_2^2] ,
\end{equation}
where the generator uses $k$-space measurements ${\bf y}$ as input. Here, the first term ensures consistency in the image domain, while the latter enforces consistency in the k-space domain. The second part of the training loss is a Wasserstein GAN loss,

\begin{align}
    {\cal L}_\textrm{WGAN}({\bm \theta}_G, {\bm \theta}_D) = \max_{||D_{{\bm \theta}_D}||_L \leq 1}\Ed_{\bf x} [D_{{\bm \theta}_D}({\bf x})] -  \Ed_{\bf z} [D_{{\bm \theta}_D}(G_{{\bm \theta}_G}({\bf z}))].
\end{align}
This equation is a generalization of Eq. \eqref{eq:minimax} with improved training stability, where the $D_{{\bm \theta}_D}$ now outputs a scalar value instead of a probability and as such is referred to as a critic instead of a discriminator, and $||D_{{\bm \theta}_D}||_L \leq 1$ indicates that it is restricted to be a Lipschitz-1 function.

 The final training loss is given as a weighted combination of these
\begin{equation}
    {\cal L}_\textrm{cycleGAN} = \gamma {\cal L}_\textrm{cycle}({\bm \theta}_G) + {\cal L}_\textrm{WGAN}({\bm \theta}_G, {\bm \theta}_D),
\end{equation}
where $\gamma$ is a weighting hyperparameter. {This approach was applied to unsupervised training of generative models for MRI reconstruction \cite{cycleGan_Jong_MRI}.}

{Variational Autoencoders (VAEs) build on the dimensionality-reducing encoder-decoder structure of autoencoders. Different from autoencoders, the encoder in a VAE learns a conditional distribution on the latent space, conditioned on the input distribution. Then, a vector is sampled from this probability distribution and fed to a decoder, which approximates the original data distribution conditioned on the latent space distribution. Hence, the latent code is learned in VAEs for a class of input images, while for conventional GANs, the latent vector amounts to random noise. VAEs are used as explicit priors in MR image reconstruction, termed deep density priors~\cite{Tezcan2019}. To approximate the data distribution of MR images, the VAE is trained on image patches that were extracted from fully-sampled reference images. After training the VAE, the learned prior model is plugged in a Bayesian formulation of a multi-coil MR reconstruction problem. Another application of VAEs in the field of MRI reconstruction is uncertainty quantification~\cite{Edupuganti2021}, where the VAE encodes the acquisition uncertainty and a Monte Carlo sampling approach is used to sample from the learned distribution and generate uncertainty maps along with the reconstructed image.}

\subsection{Algorithm unrolling and unrolled networks} \label{Sec:3d}

\begin{figure}[htp]
\includegraphics[width=\textwidth, height=\textheight, keepaspectratio]{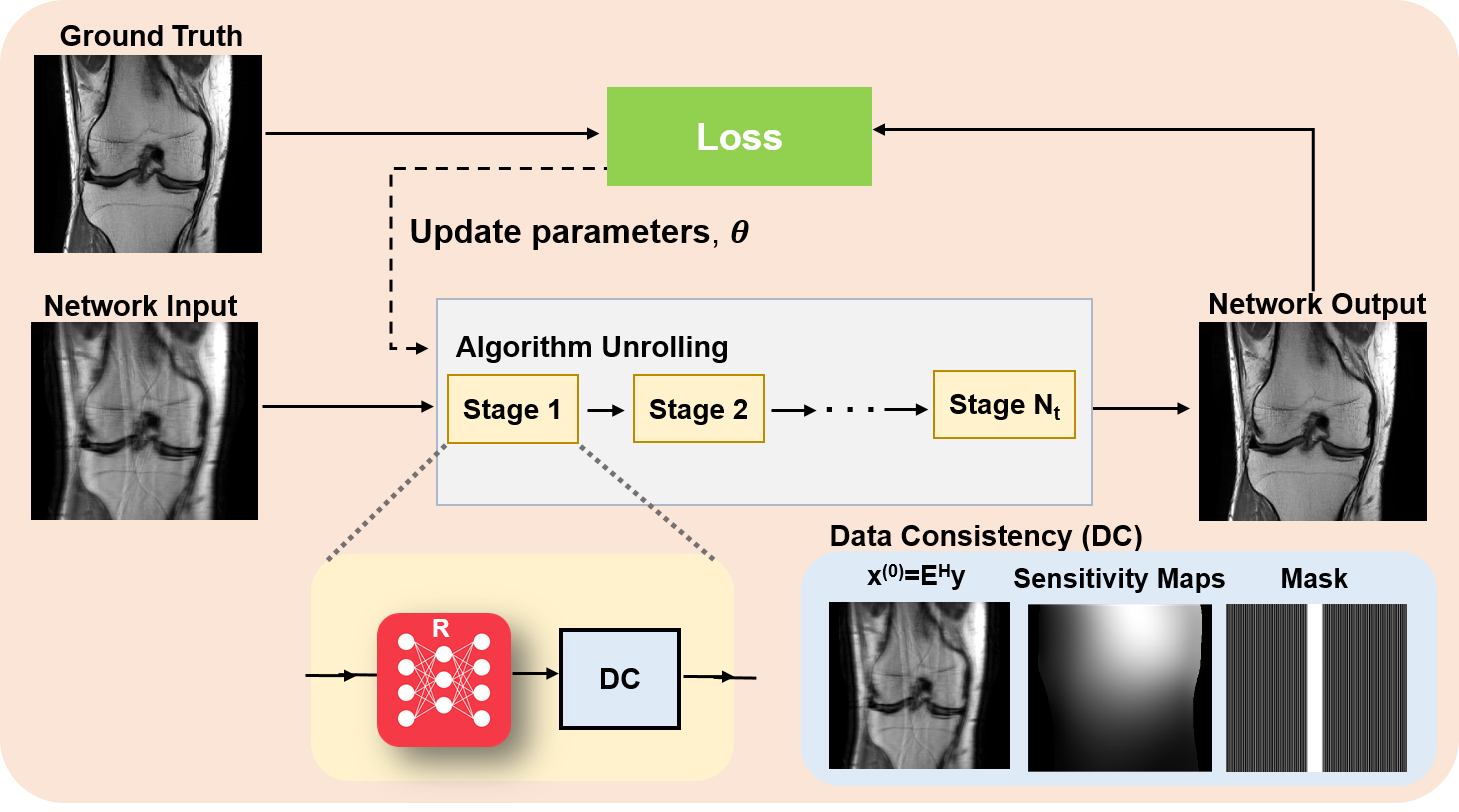}
\caption{Overview of algorithm unrolling in physics-driven deep learning methods for computational MRI. An iterative algorithm for solving Eq.~\eqref{eq:regularized_problem} is unrolled for a fixed number of iterations, and trained end-to-end using {corresponding fully-sampled/reference data}. {The red network block $R$ denotes a regularization network. This is followed by a data consistency (DC) layer. The implementation of the DC layer depends on which algorithm is used for unrolling. If a gradient descent scheme is used, the DC layer implements a gradient descent update involving the raw k-space data, and other constants that are involved in the forward model. If variable splitting based methods are used, this involves solving a problem similar to Eq. \eqref{eq:admm-dc}. The parameters can be either shared or vary over the single iterations, also termed stages.}}\label{fig:unroll}
\end{figure}

Algorithm unrolling considers the traditional iterative approaches considered in Section \ref{sec:2a} and adapts them in a manner that is amenable to learning the optimal parameters for image reconstruction \cite{Liang2020}. Traditional approaches require numerous iterations during optimization to solve the MRI reconstruction problem. Additionally, only a fixed, handcrafted regularizer is used, which do not necessarily model MR images accurately. Instead of solving a new optimization problem for each task, the whole iterative reconstruction procedure, including the image regularizer, can be learned. The original idea was proposed in the context of sparse coding \cite{Gregor2010}, but has found great use in computational imaging applications, including computational MRI. In this line of work, a conventional iterative algorithm for solving Eq.~\eqref{eq:regularized_problem} is unrolled and solved for a fixed number of iterations, as overviewed in Figure \ref{fig:unroll}. {The concept of algorithm unrolling will be introduced throughout this section.} In practice, any iterative optimization algorithm can be unrolled for solving Eq.~\eqref{eq:regularized_problem}. In the context of MRI, algorithm unrolling is based on ADMM~\cite{Yang2016} as described in Eq.~\eqref{eq:admm-dc}-\eqref{eq:admm-dual}, gradient descent schemes~\cite{Hammernik2018}, proximal gradient schemes~\cite{Aggarwal2019,Schlemper2018,Qin2019}, primal-dual methods~{\cite{Cheng2019,Ramzi2020,Ramzi2021}}, or variable splitting methods~\cite{Duan2019a,yaman_SSDU_MRM,Hosseini_JSTSP,yaman_mm_ssdu}. {Note that these algorithms contain a processing step associated with the regularization, such as the proximal operator as in Eq. \eqref{eq:pgd-updates-prox} or \eqref{eq:admm-prox}, and a data consistency step that ensures the image estimate is consistent with the acquired k-space data, such as the gradient descent step in Eq. \eqref{eq:pgd-updates-dc} or the $\ell_2$ minimization step in \eqref{eq:admm-dc}. We refer to this latter step that controls fidelity with the raw k-space data as \emph{data consistency layer} (or block).}

{We introduce the concept of unrolled networks on} Variational Networks (VNs){, which} are an example for an unrolled gradient descent scheme. In this method, the gradient descent approach in Eq.~\eqref{eq:gd-iters} is unrolled for {a fixed number of} $N_t$ steps. In VNs, the gradient {of} the regularizer $\nabla_{\bf x} \mathcal{R} ({\bf x})\vert_{{\bf x = x}^{(i-1)}}$ is derived from the Fields-of-Experts (FoE) regularizer~\cite{Roth2009}, i.e., 

$$  \mathcal{R}({\bf x}) = \sum\limits_{j=1}^{N_k} {{\bm \Phi}_j({\bf K}_j{\bf x}).}$$ 

This can be seen as a generalization of the Total Variation semi-norm for a number of $N_k$ convolution operators ${\bf K}{{_j}:\mathbb{C}^{N_x}\to \mathbb{C}^{N_x}}$ and non-linear potential functions ${\bm \Phi}{{_j}:\mathbb{C}^{N_x}\to \mathbb{R}}$. Calculating the gradient with respect to $\bf x$ yields:
\begin{align}\label{eq:FoE_grad}
    {\nabla_{\bf x} \mathcal{R}({\bf x})} = \sum\limits_{j=1}^{N_k} {\bf K}_j^H {\bm \Phi}^\prime_j({\bf K}_j{\bf x}),
\end{align}
where ${\bf \Phi}^\prime{{_j}:\mathbb{C}^{N_x}\to \mathbb{C}^{N_x}}$ denotes {the gradient vector of ${\bm \Phi}_j$ with respect to $\bf x$}
. {Plugging Eq.~\eqref{eq:FoE_grad} in Eq.~\eqref{eq:gd-iters} yields
\begin{align}
\label{eq:gd-iters-vn}
    {\bf x}^{(i)} ={\bf x}^{(i-1)}+\eta {\bf E}^H({\bf y - Ex}^{(i-1)}) - \sum\limits_{j=1}^{N_k} {\bf K}_j^H {\bm \Phi}^\prime_j({\bf K}_j{\bf x}^{(i-1)}) \quad i=0,\ldots N_t-1,
\end{align}
where $N_t$ denotes the number of cascaded stages, and one updated $(i)$ denotes a single stage. The network is said to be \emph{unrolled} for a fixed number of stages $N_i$ for training. In Eq.~\eqref{eq:gd-iters-vn} the trainable network parameters are the convolution operators ${\bf K}_j$, the activation functions ${\bm \Phi^\prime}_j$ and the weight $\eta$. The parameters can be shared over stages or varied over stages. The activation functions ${\bm \Phi^\prime}$ are modelled by a weighted combination of Gaussian radial basis functions, whose weights are learned, allowing us to approximate arbitrary activation functions.}
\begin{align}
\label{eq:gd-iters-vn}
    {\bf x}^{(i)} ={\bf x}^{(i-1)}+\eta {\bf E}^H({\bf y - Ex}^{(i-1)}) - \sum\limits_{j=1}^{N_k} {\bf K}_j^H {\bm \Phi}^\prime_j({\bf K}_j{\bf x}^{(i-1)}) \quad i=0,\ldots N_t-1,
\end{align}

VNs are characterized by the energy-based formulation of the regularizer. In the context of MRI reconstruction, the FoE model~\cite{Hammernik2018} and the Total Deep Variation model~\cite{Narnhofer2022} are proposed for energy-based regularization. In other approaches, this energy-based formulation is discarded and the gradient with respect to ${\bf x}$ is replaced by a CNN with trainable parameters ${\bm \theta}$:
\begin{align}
    \nabla \mathcal{R}({\bf x})\vert_{{\bf x = x}^{(i-1)}} = \text{CNN}_{\bm \theta}({\bf x}^{(i-1)})\label{eq:gd_regularizer_cnn}.
\end{align}

{Another line of work considers the variable splitting approach in Eq. \eqref{eq:vsqp-dc}-\eqref{eq:vsqp-prox},} 

used in data consistent CNNs \cite{Schlemper2018} and MoDL \cite{Aggarwal2019},  which  again replace the gradient with respect to ${\bf x}$ by a CNN with trainable parameters ${\bm \theta}$ as in Eq. \eqref{eq:gd_regularizer_cnn}. This leads to the following scheme:
\begin{subequations}
\begin{align}
{\bf x}^{(i)} &= \big({\bf E}^H{\bf E} + \eta {\bf I}\big)^{-1} \big({\bf E}^H{\bf y} + \eta {\bf z}^{(i)}), \label{eq:pgd-unroll-dc-alt}\\
{\bf z}^{(i)} &= \text{CNN}_{\bm \theta}({\bf x}^{(i-1)}) \label{eq:pgd-unroll-cnn-alt}
\end{align}
\end{subequations}
where $\eta$ is an additional learnable parameter. {Eq.~\eqref{eq:pgd-unroll-dc-alt} can be solved directly via matrix inversion for single-coil datasets \cite{Schlemper2018}, or using an iterative optimization approach based on conjugate gradient (CG) for the more commonly used multi-coil setup \cite{Aggarwal2019}, where matrix inversion is computationally infeasible.} Note in this case, the CG algorithm itself has to be unrolled for a fixed number of iterations for easy back-propagation through the whole network. Once again, the CNN in Eq.~\eqref{eq:pgd-unroll-cnn-alt} can be any kind of regularization network, as the idea is agnostic to the particulars of the CNN that is used in this step.

Proximal gradient descent unrolling, which utilizes Eq. \eqref{eq:pgd-updates-dc}-\eqref{eq:pgd-updates-prox}, leads to the replacement of the proximal operator of ${\cal R}(\cdot)$ by a CNN with trainable parameters ${\bm \theta}$, leading to:  
\begin{subequations}
\begin{align}
{%
{\bf z}^{(i)}} &= {{\bf x}^{(i-i)} + \eta {\bf E}^H({\bf y-Ex}^{(i-i)}).}\label{eq:pgd-unroll-dc}\\
{%
{\bf x}^{(i)}} &= {\text{CNN}_{\bm \theta}({\bf z}^{(i)})}\label{eq:pgd-unroll-cnn}
\end{align}
\end{subequations}
This method was utilized in \cite{Mardani2018Neurips}. This approach leads to a less memory intensive data consistency step in Eq. \eqref{eq:pgd-unroll-dc} compared to  Eq. \eqref{eq:pgd-unroll-dc-alt}, especially for multi-coil datasets, but its performance lags the CG approach used in MoDL \cite{Hosseini_JSTSP}. 

{In the context of learned unrolled schemes, classical multi-layer CNNs~\cite{Schlemper2018,Aggarwal2019} or multi-scale regularizers such as UNET~\cite{sriram2020endtoend}, Down-Up Networks~\cite{Hammernik2021}, multi-level wavelet CNNs~\cite{Ramzi2021,Liu2018} are commonly used.} {Also, the parameters of these networks can be either shared, e.g.~\cite{Aggarwal2019}, or varied, e.g.~\cite{Hammernik2018}, over the stages}. 

However, similar performance has been achieved with both gradient descent and variable splitting-type algorithms as reported in the first fastMRI reconstruction challenge~\cite{Knoll2019_fastMRI}. The differences reported in the context of the second fastMRI reconstruction challenge focus more on managing different coil sensitivities and regularization networks~\cite{fastmri_2ndChallenge}.

\subsubsection{Training unrolled networks} \label{sec:3d1}

The output of the unrolled network depends on the variables in both the regularization {network} and data consistency 
{layers}, and can be represented with a function $f_\textrm{unroll}({\bf y}, {\bf E}; \{{\bm \theta}_i, \eta_i\}_{i=1}^{N_t}). $
{For} the most generalized representation, we allow the regularizer CNN parameters ${\bm \theta}$ and the data consistency parameters ${\eta}$ to vary across the unrolled iterations {(cascaded stages). However, as noted earlier, the parameters can also be shared between stages}. While for ease of notation, we have referred to the multi-coil operator as ${\bf E}$, this operator implicitly includes the sub-sampling mask $\Omega$. For the following, we will make this dependence explicit, and use ${\bf E}_{\Omega}$ and ${\bf y}_{\Omega}$ for the multi-coil operator and the measured k-space data, respectively.

The standard learning strategy for unrolled networks is to train them end-to-end, using the full network that has been unrolled for $N_t$ steps. For end-to-end training of unrolled networks, the most commonly used paradigm relies on supervised learning, where a database of fully-sampled measurements/ground-truth images as a reference. Given a database of pairs of input and reference data, the supervised learning loss function can be written as
\vspace{-0.02cm}
\begin{equation}
    \min_{\{{\bm \theta}_i, \eta_i\}_{i=1}^{N_t}} \frac1N \sum_{n=1}^{N} \mathcal{L}( {\bf x}_{\textrm{ref}}^n, \:f({\bf y}_{\Omega}^n, {\bf E}_{\Omega}^n;  \{{\bm \theta}_i, \eta_i\}_{i=1}^{N_t})),
    \vspace{-0.02cm}
\end{equation}
where ${\bm \theta}$ represents the network parameters, $N$ is the number of samples in the training database, $\mathcal{L}(\cdot, \cdot)$ is a loss function characterizing the difference between network output and referenced data, ${\bf x}_{\textrm{ref}}^n$ denotes the ground-truth image for subject $n$. The domain for the loss function can be image, k-space or a mixture of them. Numerous loss functions such as $\ell_1$, $\ell_2$, adversarial and perceptual losses have been used in supervised deep learning approaches \cite{Knoll_SPM}.

However, in many applications, fully-sampled reference data may be impossible to acquire, for instance due to organ motion or signal decay, or may be impractical due to excessively long scan times. In these cases, self-supervised learning enables training of neural networks without fully-sampled data by generating {training data} from the sub-sampled measurements themselves \cite{akcakaya_unsupervised_SPM}. One of the first works in this area, self-supervised learning via data undersampling (SSDU) \cite{yaman_SSDU_MRM}, partitions the acquired measurements $\Omega$, for each scan into two disjoint sets, $\Theta$ and $\Lambda$. One of these sets, $\Theta$, is used during training to enforce data consistency within the network, while the other set, $\Lambda$, remains unseen by the unrolled network and is used to define the loss function in k-space. Hence, SSDU performs end-to-end training by minimizing the following self-supervised loss: 
\begin{equation}
    \min_{\bm \theta} \frac1N \sum_{n=1}^{N} \mathcal{L}\Big({\bf y}_{\Lambda}^n, \: {\bf E}_{\Lambda}^n \big(f({\bf y}_{\Theta}^n, {\bf E}_{\Theta}^n; {\bm \theta}) \big) \Big),
    \vspace{-0.02cm}
    \label{eq:ssdu}
\end{equation}
where  the network output is transformed back to k-space by applying the encoding operator ${\bf E}_{\Lambda}^n$ at unseen locations in training. Thus, the self-supervised loss function measures the reconstruction quality of the model by characterizing the discrepancy between the unseen acquired measurements and network output measurements at the corresponding locations. Once the network is trained, the reconstruction for unseen test data is performed by using all acquired measurements $\Omega$. In another line of work \cite{ensure}, Stein's unbiased risk estimate of mean square error (MSE) is leveraged to enable unsupervised {training of neural networks for} MRI reconstruction. In particular, the loss function obtained from an ensemble of images, each acquired by employing different undersampling operator, has been shown to be an unbiased estimator for MSE.  

Finally, like generative models based on DIP, there has been interest in training unrolled networks on single datasets without a database. In this setting, the number of trainable parameters is usually larger than the number of pixels/k-space measurements, and training may lead to overfitting. Recent work in this area has tackled this challenge by developing a zero-shot self-supervised learning \cite{zs_ssl} approach that includes a third additional partition, which is used to monitor a self-validation loss in addition to the previous self-supervision setup. This self-validation loss starts to increase once overfitting is observed. {Once the model training is stopped, the final reconstruction is calculated by using the network parameters from the stopping epoch, while using all acquired measurements.}

\subsubsection{Memory challenges of unrolled networks {and deep equilibrium networks}}
A major challenge for training unrolled networks is their large memory footprint. When an algorithm is unrolled for $N_t$ iterations, a straightforward implementation involves the storage of $N_t$ CNNs, along with $N_t$ DC operations in GPU memory. The latter itself can have a large footprint, when a CG-type approach is used \cite{Aggarwal2019}. This creates challenges for training unrolled networks for large-scale or multi-dimensional datasets, especially since deeper networks tend to lead to better performance \cite{Pezzotti2020}. Recently, this was tackled with the development of memory-efficient learning schemes \cite{kellman_TCI_2020}. In memory-efficient learning, intermediate outputs from each unrolled iteration are stored on host memory during forward pass, and backpropagation gradients are computed using this intermediate data and gradients from the preceeding step. Thus, this approach conceptually supports as many unrolling steps as desired, with the drawback of additional data transfer between GPU and the host memory.

Another alternative for handling the large memory footprint of unrolled networks is deep equilibrium networks \cite{deep_equilibrium_2021}. {Unrolled networks that share learnable weights across stages show competitive performance \cite{Aggarwal2019}, while each stage can be represented with a single function leading to a compact representation. Unrolled networks execute this function for a finite number of steps $N_t$, whereas deep equilibrium networks characterize the limit as the $N_t \to \infty.$ Provided this limit exists, it corresponds to the solution of the fixed point equation for an operator corresponding to a single 
{stage}}. This approach leads to two advantages for training. First, only one {stage} has to be stored during training, leading to a smaller memory usage. Second, {the convergence behavior} for different values of $N_t$ during inference is more well-behaved compared to unrolled networks, which are designed to achieve maximal performance for a specific value of $N_t$. On the other hand, deep equilibrium networks are run until convergence and do not have fixed inference time unlike unrolled networks, which may not be ideal in clinical applications. {Furthermore, these approaches require large training time due to the Jacobian inversion during gradient computation. Recently, this challenge has been tackled by approximating this inverse Jacobian using quasi-Newton matrices from the forward pass with substantial savings in computation cost \cite{shine}.}

\section{State-of-the-art in MRI practice and domain-specific challenges}

\subsection{Real vs complex building blocks}

As complex-valued data is used in computational MRI, this has to be considered in the network processing pipeline, not only during data consistency, but also in the network blocks itself. Two processing modes are possible: 1) Real/Imaginary or magnitude/phase are considered in two input channels stacked via the feature dimension, 2) Complex-valued operations are performed on complex-valued tensors. While the former allows us to use real-valued operations, the complex{-valued} relationship between real and imaginary parts is lost. Complex-valued operations maintain the complex nature of the data, but some operations require twice the amount of trainable parameters. For example in complex-valued convolutions, a real and imaginary filter kernel needs to be learned. Additionally, the number of multiplications doubles compared to real-valued processing. If complex-valued layers and tensors are involved, complex back-propagation following Wirtinger calculus has to be considered \cite{Virtue2018, Trabelsi2017} which is supported in most recent frameworks (Tensorflow $\geq$ v1.0, PyTorch $\geq$ v1.10). An overview of the most common layer operations together with their complex-valued Wirtinger derivatives is shown in Table \ref{tab:WirtingerCalculus}. In the context of MRI reconstruction, complex-valued processing is conducted in both ways.

\begin{table}[ht!]
  \centering
  \title{Wirtinger derivatives}
  \caption{Overview of important functions along with their pair of Wirtinger derivatives.}
    \begin{tabular}{cccc}
    \toprule[2pt]
    Function & $f(\x)$ & $\fracdd{f}{\x}$ & $\fracdd{f}{\x^H}$\\ \midrule[1pt]
    Magnitude & $\abs \x = \left(\x^H\x\right)^{0.5}$ & $\frac{\x^H}{2f(\x)}$ & $\frac{\x}{2f(\x)}$\\
    Phase & $-\text{i} \log \frac{\x}{\abs{\x}}$ & $-\frac{\text{i}}{2\x}$ & $\frac{\text{i}}{2\x^H}$ \\
    Real Component & $\frac{1}{2}\left(\x + \x^H\right)$ & $\frac{1}{2}$ & $\frac{1}{2}$\\
    Imaginary Component & $\frac{1}{2\text{i}}\left(\x + \x^H\right)$ & $\frac{1}{2\text{i}}$ & $\frac{\text{i}}{2}$\\ \midrule[0.5pt]
    Normalization & $\frac{\x}{\left(\x^H\x\right)^{0.5}}$ & $\frac{1}{2 \left(\x^H\x\right)^{0.5}}$ & $-\frac{z^2}{2 \left(\x^H\x\right)^{1.5}}$ \\
    Scalar product & $\mathbf{w}^H \x$ & $\mathbf{w}^H$ & 0 \\ \midrule[0.5pt]
    Max Pooling & $\x_n$, $n=\arg\max_k \abs{\x_k}$ &  $\begin{cases}1 & \text{if } n=\arg\max_k \abs{\x_k} \\ 0 & \text{else}  \end{cases}$& 0\\
    Dropout & $\begin{cases}\frac{1}{p}\x_n & \text{if } n \in \Omega \\ 0 & \text{else} \end{cases}$ & $\begin{cases}\frac{1}{p} & \text{if } n \in \Omega \\ 0 & \text{else} \end{cases}$ & 0\\ \midrule[0.5pt]
    Separable activation\\ (ReLU, Sigmoid, ...) & $f(\Real{\x})+ \text{i} f(\Imag{\x})$ & $\frac{1}{2}\left(\fracdd{f}{\x}(\Real{\x}) + \fracdd{f}{\x}(\Imag{\x}) \right)$ & $\frac{1}{2}\left(\fracdd{f}{\x}(\Real{\x}) - \fracdd{f}{\x}(\Imag{\x}) \right)$ \\
    Cardioid~\cite{Virtue2018} & $\frac{1}{2}\left(1 + \cos(\angle \x) \right)\x$ & $\frac{1}{2} + \frac{1}{2}\cos(\angle\x) + \frac{\text{i}}{4}\sin(\angle\x)$& $-\frac{\text{i}}{4}\sin(\angle \x)\frac{\x}{\x^H}$ \\
    Complex sigmoid & $\frac{1}{1+ e^{-{\x}}}$ & $\frac{e^{-{\x}}}{\left(1+ e^{-{\x}}\right)^2}$ & 0 \\
 \bottomrule[2pt]
    \end{tabular}%
  \label{tab:WirtingerCalculus}
\end{table}

\subsubsection{Convolution}
The discrete convolution maps the $\nfin$ input feature channels to $\nfout$ output feature channels of an image $\mathbf{x} \in \mathbb{K}^n$ with filter kernels $\mathbf{k}_{i,j} \in \mathbb{K}^k$ {of kernel size $k$} via
\begin{align}
    \hat{\x}_j = \sum\limits_{i=1}^{\nfin} \x_i \ast \mathbf{k}_{i,j} \quad j=1,\ldots,\nfout, 
    \label{eq:GeneralConvolution}
\end{align}
where the subscripts denote the feature channels. {Convolutions can be performed along multiple dimensions.} For real-valued convolutions it is $\mathbb{K} = \mathbb{R}$. For complex convolutions, $\mathbb{K} = \mathbb{C}$, the convolution operation is extended to
\begin{align}
    \x_i \ast \mathbf{k}_{i,j} &= (\Real{\x_i}\ast\Real{\mathbf{k}_{i,j}} - \Imag{\x_i}\ast\Imag{\mathbf{k}_{i,j}}) \nonumber \\
    &\phantom{= }+ \text{i}\cdot (\Imag{\x_i}\ast\Real{\mathbf{k}_{i,j}} + \Real{\x_i}\ast\Imag{\mathbf{k}_{i,j}})
\end{align}

\subsubsection{Activation}
While convolution functions operate in a local neighborhood of a pixel, activation functions operate in a pixel-wise way. When applying non-linear activation functions $\phi$ to complex values, the impact on magnitude and phase information needs to be considered. {The reader is referred to \cite{Cole2021, Virtue2018} for some comparative work.} One possibility is to apply the activation function to the real and imaginary part separately as separable activations, however, the natural correlation between real and imaginary channels are not considered in this case. {When using separable ReLUs}, phase information is mapped to the first quadrant, i.e., the interval $[0, \frac{\pi}{2}]${, as all negative real and imaginary parts are set to zero and only the positive parts are kept}.
Alternative approaches have been proposed that retain phase information, for example siglog{, defined as
\begin{align*}
    \phi_{\text{siglog}}=\frac{\bf x}{1 + \abs{\x}}.
\end{align*}}
As another option, the phase information can be fixed and only the magnitude information is altered by the activation. An example therefor{e} is the \emph{ModReLU}
\begin{align}
    \phi_\text{ModReLU}(\x) = \max (0, \abs{\x} + \beta) \frac{\x}{\abs{\x}}
\end{align}
where $\beta$ is the bias that is trainable. A new complex activation function called \emph{Cardioid} was also proposed for MRI processing \cite{Virtue2018}
\begin{align}
    \phi_{\text{Cardioid}}(\x) = \frac{1}{2}\left(1 + \cos(\angle \x + \beta) \right) \x
\end{align}
The complex cardioid can be seen as a generalization of ReLU activation functions to the complex plane. Compared to other complex activation functions, the complex cardioid acts on the input phase rather than the input magnitude. A bias $\beta$ can be additionally learned.

{The specific choice among these complex activation functions is application-dependent. For phase sensitive applications, such as water-fat imaging and phase contrast imaging, it was shown that complex networks outperformed the real-valued networks, with the separable ReLUs performing best \cite{Cole2021}, whereas for MR fingerprinting, the complex cardioid outperformed other activations functions \cite{Virtue2018}. }

\subsubsection{Normalization}
Adding normalization layers (batch, instance or layer/group normalization) directly after convolution layers are a common way to enable faster and more stable training of networks. Statistics are estimated from the input and used to re-parametrize the input. Complex-valued normalization layers require to estimate the normalization via the covariance matrix~\cite{Trabelsi2017} and are straight-forward to implement. The subsequent layers are less tolerant to changes in previous layers. The selection of the normalization layer is task dependent. Although normalization layers are often important to train a network, they might lead to unwanted artifacts for image restoration tasks \cite{Wang2018a}. 

\subsubsection{Pooling}
Pooling layers are used as down-sampling operation to reduce the spatial resolution in the image and to introduce approximate invariance to small translations.  Small patches are analyzed in the individual features maps to keep important information about extracted features. Common pooling layers are \emph{Average Pooling} and \emph{Max Pooling}. For complex-valued images, the maximum operation does not exist. Instead, the pooling layer is modified such that it keeps values with, e.g., the maximum magnitude response.

\begin{figure}[htp]
    \centering
    \includegraphics[width=0.9\textwidth]{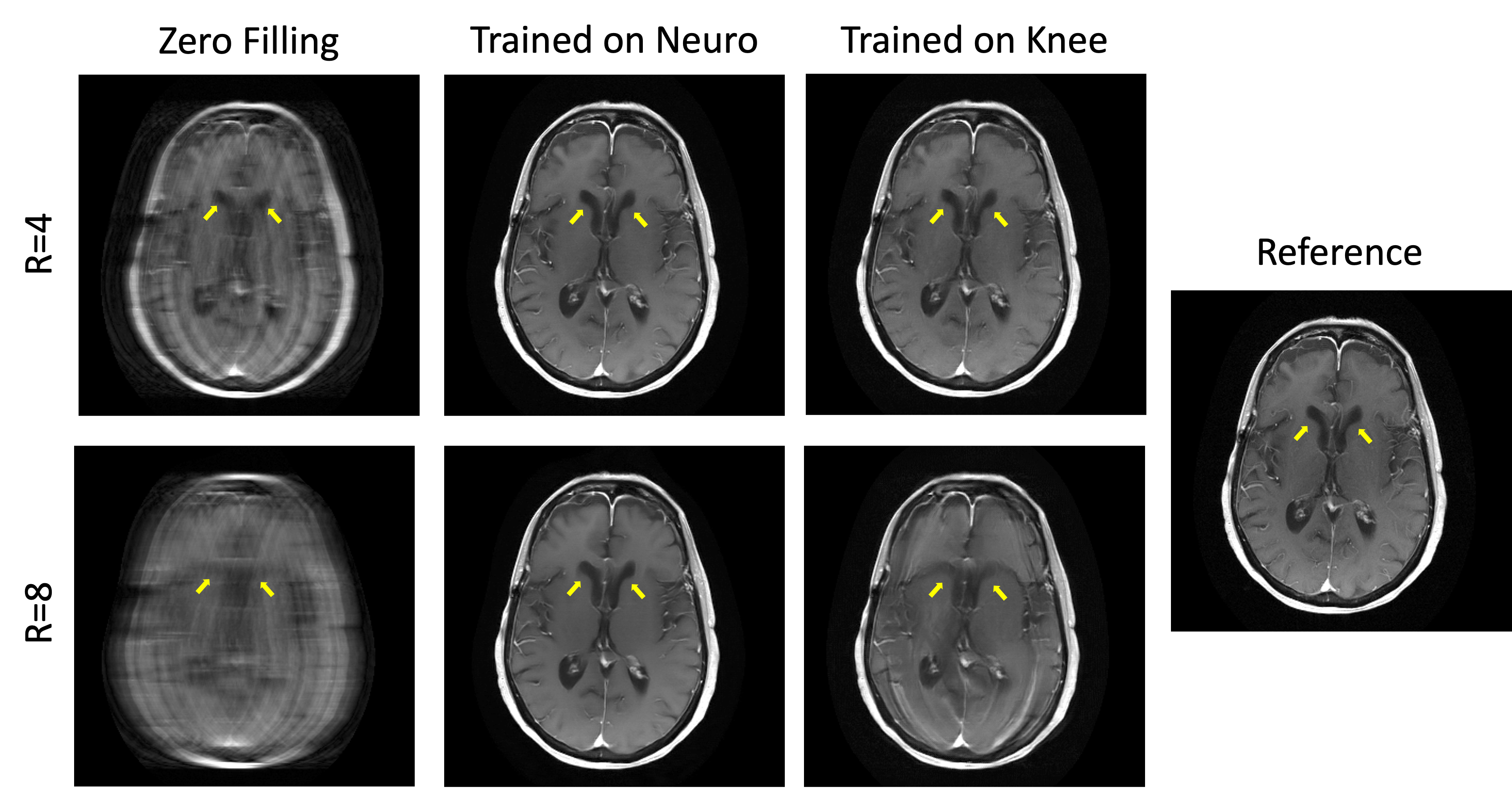}
    \caption{Down-Up Networks combined with a proximal mapping layer for data consistency~\cite{Hammernik2021}, trained with different data configuration. While the reconstruction performance generalizes well independent of the type of training data for R=4, the ventricles of the brain change here when trained with the wrong data, i.e., knee data for R=8.}
    \label{fig:knee_neuro}
\end{figure}

\subsection{Canonical MRI reconstruction with the linear forward model}

Physics-driven MRI deep learning methods have become the most popular approach in computational MRI due to their improved robustness, especially for the accelerated MRI problem that relies on the linear forward model in Eq. \eqref{eq:forwardbasic}. Such methods have been the top performers in community-wide reconstruction challenges, such as the fastMRI challenge~\cite{Knoll2020a,Muckley2021}, {for Cartesian sampling. The success for physics-driven learning for MRI reconstruction is not limited to the Cartesian sampling pattern. Promising results were also shown for non-Cartesian sampling schemes \cite{rare, Ramzi2022, Zhang2022a}, including works that learn non-Cartesian trajectories and reconstruction jointly \cite{bjork2022, Chaithya2022a}.}  %Indeed, the difference between the state-of-the-art algorithms for the fastMRI challenge is hardly visible for the different undersampling factors.
These algorithms have in common that data consistency is included, and expressive regularization networks are used. Imaged pathologies do not {need to} be included in the training dataset { as long as enough k-space data is available to guide the reconstruction to recover the pathology encoded in k-space. Additionally it was shown that the pathologies may appear or disappear depending on the selection of the undersampling pattern for a given number of sampled k-space lines~\cite{Johnson2021}. Theoretical analysis, reader studies and uncertainty quantification as proposed in~\cite{Narnhofer2019} are tools that might support us to identify the clinically possible acceleration limit.} 
%as long as the acceleration factors is not pushed too far and the pathologies are encoded in the k-space~\cite{Johnson2021}.

However, even physics-driven deep learning methods face some challenges for accelerated MRI. The impact of domain shift, i.e., training and testing on different data was studied in~\cite{Hammernik2021}, for different acceleration factors. All training and evaluation is based on the fastMRI knee and neuro datasets~\cite{Knoll2020a}. While for acceleration 4, the proposed Down-Up networks with varying data consistency layers generalize well for both anatomies, the type and amount of training data becomes more critical for acceleration factor 8. Since fewer data is available for data consistency at this acceleration, the networks start to reconstruct anatomical structures that are not real. When trained on a subset of knee data and applied to neuro data, the ventricles start resembling knee structures, for acceleration 8 as depicted in Figure~\ref{fig:knee_neuro}. 

\begin{figure}[htp]
    \centering
    \includegraphics[width=\textwidth]{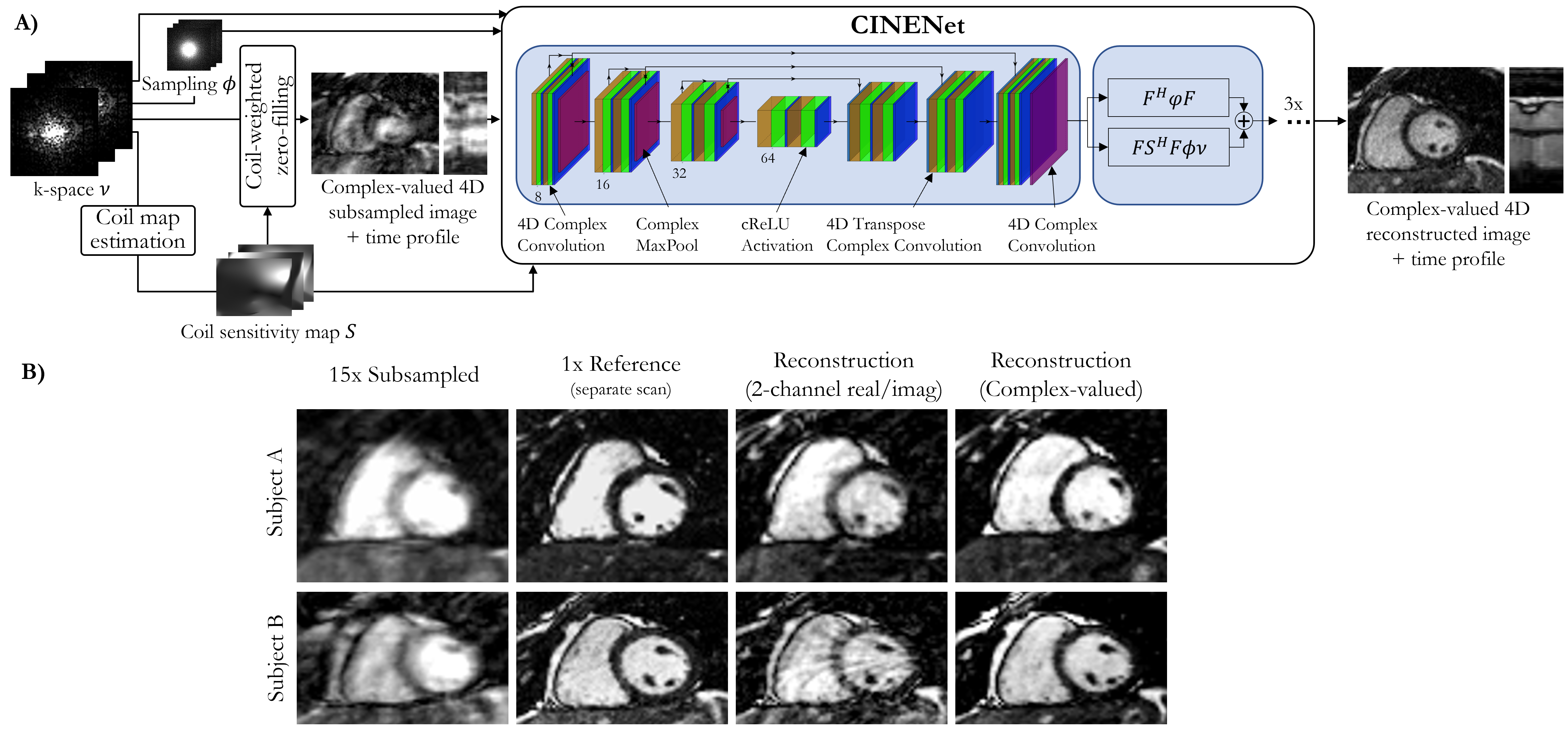}
    \caption{{A) CINENet combines data consistency layers and a UNet architecture with complex-valued building blocks for convolution, activation, normalization and pooling layers. To process the 3D+$t$ data, convolutions are split into 3D spatial convolutions and 1D temporal convolutions. B) Impact of complex-valued operations over 2-channel (real/imaginary) processing in two subjects for a prospectively undersampled 3D cardiac CINE (3D+$t$) acquisition with R=15. A separate fully-sampled (R=1) reference scan is obtained for comparison.}}
    \label{fig:cinenet}
\end{figure}

All previously mentioned approaches consider the complex-valued MR images as images with two real-valued feature channels. CINENet~\cite{CineNet} combined both data consistency layers with complex-valued building blocks as depicted in Figure~\ref{fig:cinenet}, for dynamic 3D (3D+$t$) data. These complex-valued building blocks include convolutions, activations, pooling, and normalization layers. To process the 3D+$t$ in the regularization network, convolution operations are split into 3D spatial convolutions, followed by 1D temporal convolutions.

\begin{figure}[htp]
    \centering
    \includegraphics[width=\textwidth]{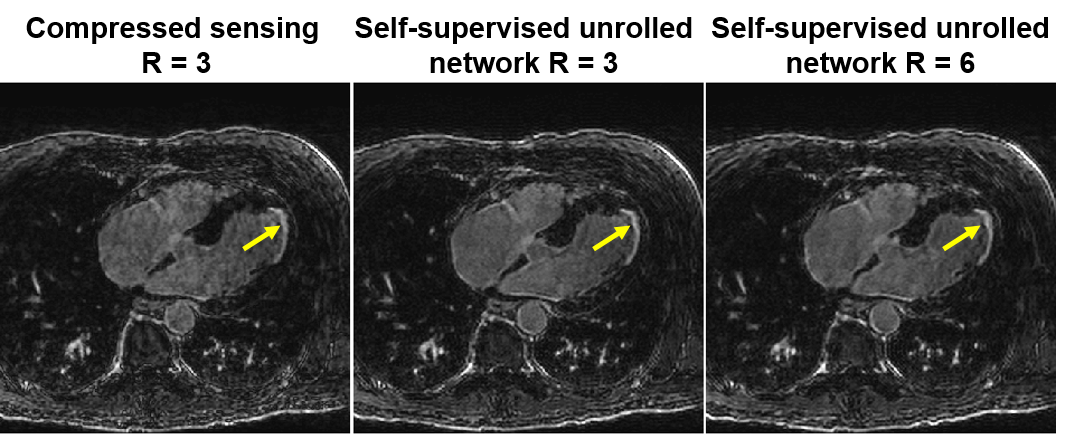}
    \caption{Reconstruction results from a high-resolution late gadolinium enhancement acquisition on a cardiac patient (arrows: scar areas). This scan cannot be fully-sampled due to contrast-related scan time constraints. Unrolled networks can be trained in a self-supervised manner \cite{yaman_SSDU_MRM}, leading to reconstructions that outperform current clinically used approaches, such as compressed sensing, and allowing acceleration rates twice as fast \cite{YamanLGE}.}
    \label{fig:lge-ssdu}
\end{figure}

Finally, as aforementioned the need for fully-sampled data for training had hindered the use of deep learning reconstructions for certain applications. Thus, alternative methods have been explored. Dynamic contrast-enhanced MRI (DCE-MRI) represents one such challenging acquisition, where $k$-space data is acquired continuously while contrast agent been injected to the patient. The dynamic distribution of the contrast agent causes the image contrast dynamics, hence, both the $k$-space and image are time-series. In this setting, a variational network was trained on simulated data with radial $k$-space trajectories, as fully sampled dataset with both high spatial and temporal resolution was infeasible to acquire \cite{Huang2021}. With the recent advances outlined in Section \ref{sec:3d1}, training in such scenarios can also be done in with more realistic datasets without resorting to simulations. For instance, in another contrast-based cardiac acquisition, called late gadolinium enhancement imaging, unrolled networks have been trained using prospectively accelerated acquisitions without {fully-sampled reference data} \cite{YamanLGE}, and were shown to improve on clinically used compressed sensing methods, doubling the achievable acceleration rates, as depicted in Figure \ref{fig:lge-ssdu}.

\subsection{Inverse problems in MRI with non-linear forward models} 

Recently, deep learning models have been proposed to address the computationally demanding task of non-linear inverse problems in MRI. 
A neural network $f_{\textrm{NM}, {\bm \theta}}: {\mathbb C}^{m \cdot n_c} \to {\mathbb C}^{n_v}$, parametrized by ${\bm \theta}${, which maps the acquired data ${\bf y}$ to the unknown parameters ${\bf v}$ (e.g. magnetization and relaxation maps)} is learned either in a supervised setup \cite{liu2019mantis}:
{
\begin{equation}
    \arg\min\limits_{\bm \theta} {\mathbb E}_{\bf y} \: \|{\bf E}{\cal M}(f_{\textrm{NM}, {\bm \theta}}({\bf y})) - {\bf y}\|_2^2 + \lambda ||f_{\textrm{NM}, {\bm \theta}}({\bf y})-{\bf v}_{\textrm{ref}}||_2^2  \label{eq:supervised-nonlinear}
\end{equation}
}
or in a self-supervised setup \cite{Liu2021}
{
\begin{equation}
    \arg\min\limits_{\bm \theta} {\mathbb E}_{\bf y} \: \|{\bf E}{\cal M}(f_{\textrm{NM}, {\bm \theta}}({\bf y})) - {\bf y}\|_2^2 + {\cal R}(f_{\textrm{NM}, {\bm \theta}}({\bf y})),  \label{eq:unsupervised-nonlinear}
\end{equation}
}
where figure{${\cal R}(\cdot)$} {is a} conventional {regularizer} that {is} not based on reference data, such as spatial total variation. In the following, we will expand on some applications for which non-linear forward models are beneficial.

\subsubsection{Relaxivity mapping}
MRI allows for quantitative measurements of inherent tissue parameters (T$_1$, T$_2$, T$_2^*$, T$_{1\rho}$), which is often referred to as relaxivity or quantitative mapping. In recent years, research developments have contributed towards the goal of retrieving multiple parametric maps from a single scan \cite{seiberlich2020quantitative}. A model-based reconstruction in these cases eliminates the need for reconstructing individual images along the relaxivity curve {(data sampled at different time-points along the T$_1$/T$_2$/T$_2^*$/T$_{1\rho}$ relaxation based on Eq. \eqref{eq:nonlinearModelKspace} after excitation with appropriate preparatory pulses).} 
However, model-based reconstruction methods have prolonged reconstruction times compared to reconstruction of individual images followed by a parametric fitting, which hinders their clinical translation \cite{seiberlich2020quantitative}. 
Deep learning models have been proposed to enable fast inference and shifting time-demanding workloads to the offline training procedure, showing feasibility in a number of quantitative mapping applications \cite{liu2019mantis,Virtue2018}. In this setting, physics information{, arising from the underlying known forward model (Eq. \eqref{eq:forwardbasic} or \eqref{eq:forwardnonlinear}),} has primarily been incorporated to the loss function during training, similar to Eqs. \eqref{eq:supervised-nonlinear} and \eqref{eq:unsupervised-nonlinear} \cite{liu2019mantis,Liu2021,KamilovR2}.

\subsubsection{Susceptibility mapping}
Physics-driven deep learning methods have also been studied in the context of quantitative susceptibility mapping. First works incorporated the physical principles of the dipole inversion model that describes the susceptibility-phase relationship into the loss function during neural network training \cite{yoon2018quantitative}. Later works optimized the parameters in an unrolled gradient descent algorithm for {non-linear} dipole inversion \cite{polak2019nonlinear, Feng2021}. More recently, the idea of fine-tuning pre-trained network weights on a scan-specific basis using the physics model was proposed \cite{Zhang2020fine}, similar to the loss function in Eq. \eqref{eq:unsupervised-nonlinear} without the additional regularizers.

\subsubsection{Flow, perfusion and contrast-enhanced MRI}
Flow imaging is used to measure blood velocities in the circulatory system, but requires considerable acquisition time. Thus, physics-driven deep learning methods have been proposed for improving flow imaging. These methods enforce fidelity to the physical model in the losses to a known ground-truth \cite{ferdian20204dflownet}, or incorporate the physics of fluid flow by solving the partial differential equations via automatic differentiation in backpropagation \cite{Fathi2020}. Perfusion techniques provide ways to model blood flow often via the use of a kinetic model {(describing the perfusion of the tissue) \cite{Buxton1998}}, but suffer from low-resolution or low signal-to-noise ratio acquisitions. Physics-driven methods incorporating the kinetic model to the loss function has been proposed \cite{ulas2018deepasl}. In a similar vein, estimation of pharmacokinetic parameters {(describing tracer kinetics of the tissue) \cite{Tofts1999}} from dynamic contrast-enhanced MRI by residual learning using the physical forward model \cite{ulas2018direct} was also proposed for faster parameter inference.

\subsubsection{Motion}
\label{sec:nonlinearMotion}

Acquisitions under physiological and patient motion require methods for handling motion in order to avoid aliasing or blurring of the imaged anatomy. In addition to various prospective motion triggering, gating or correction methods, motion can be retrospectively modeled into the forward model and can thus be considered inside a motion-compensated/corrected reconstruction \cite{Odille2008a,batchelor2005matrix}. These methods perform two fundamental operations: image registration and image reconstruction. Hence, they require reliable motion-resolved images from which the motion can be estimated. Motion field estimation can be controlled or supported by external motion surrogate signals \cite{Odille2008a} or initial motion field estimates \cite{batchelor2005matrix}. 

\begin{figure}[htb]
    \centering
    \includegraphics[width=\textwidth]{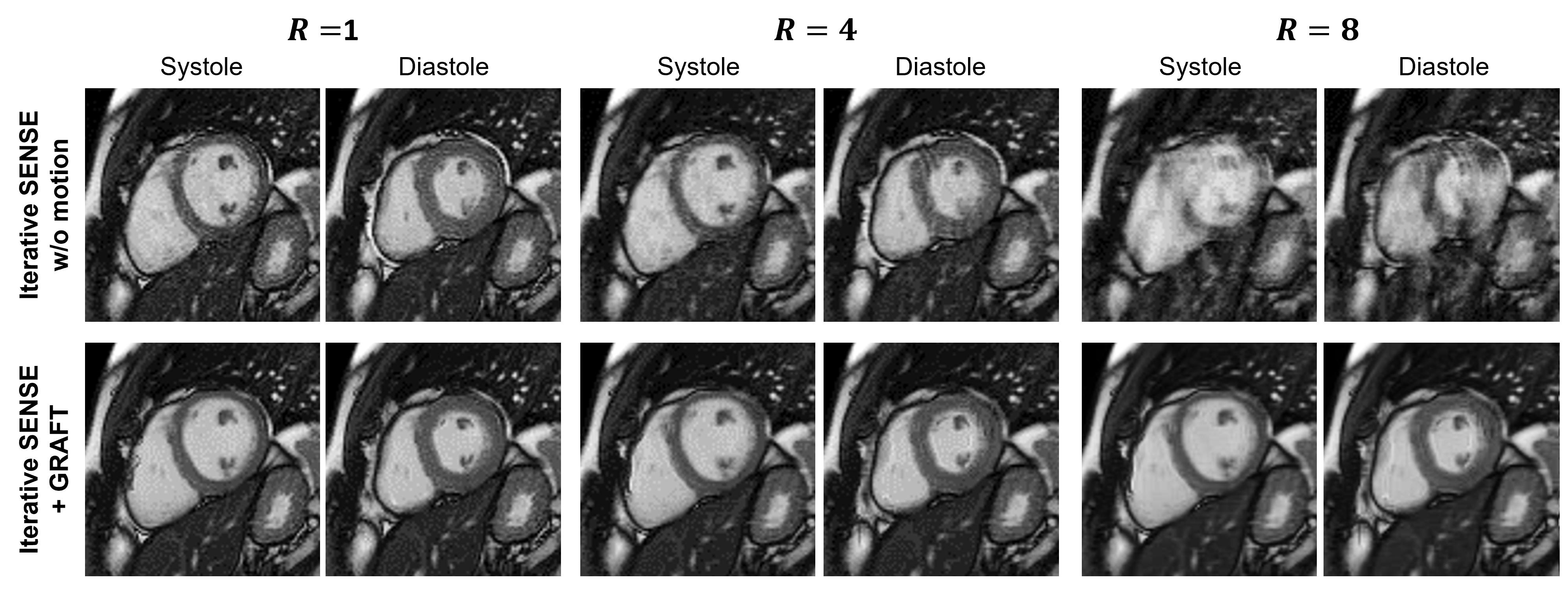}
    \caption{Motion-Compensated reconstruction. A motion estimation network (GRAFT) is embedded in the reconstruction procedure~\cite{Hammernik2021asilomar}. The motion-compensated reconstruction outperforms iterative SENSE without motion compensation. If motion compensation is not performed, undersampling artifacts are substantially present. Systole and diastole frames are depicted for R=1, R=4 and R=8.}
    \label{fig:moco}
\end{figure}

While deep learning allows for efficient motion estimation, only few works embed motion estimation in image reconstruction. Among these, LAPNet formulates non-rigid registration directly in k-space \cite{KuestnerAPSIPA2020, lapnet}, inspired by the optical flow formulation. The estimated motion fields are then used to enhance the data consistency and exploit the information of all motion resolved states to reconstruct images of the body trunk. In the context of coronary MRI, a motion-informed MoDL network was proposed \cite{Qi2021}, using diffeomorphic motion fields estimated from the zero-filled images using a UNet and subsequent scaling and squaring layer. These motion fields are then embedded into the data consistency layer, solved via the conjugate gradient algorithm as in MoDL. The network is unrolled for 3 iterations, with intermittent denoising networks. The full model is trained using a reconstruction loss and a motion estimation loss. For dynamic MR images, the whole temporal information was exploited by embedding motion-estimation UNets directly in the data consistency layers of an unrolled network architecture \cite{Seegolam2019}. Hence, both reconstruction and motion estimation improve as the motion-estimation networks rely on the reconstructions of the previous unrolled iteration. Further approaches achieved motion correction by rejecting motion-affected k-space lines~\cite{Oksuz2020} or subspace-constrained regularization~\cite{biswas2019dynamic}. Inspired by~{\cite{Cruz2021singleheartbeat,batchelor2005matrix}}, {warping with a motion field} is embedded in {the} {forward operator} for Cartesian cine imaging{, where the motion fields are estimated by a neural network} ~\cite{Hammernik2021asilomar}. An example reconstruction results of the systole and diastole for accelerations R=4 and R=8 is depicted in Figure~\ref{fig:moco}.

\section{Discussion}
\subsection{Issues and open problems}

Deep learning has dominated research in computational MRI during the last few years, and while there are still a number of open questions and issues, both on the basic science and on the translational front, they evolved as the developments are going on in the field. During the early stages of the developments, access to raw MRI k-space training data was a major limiting factor hat held the field back. The availability of open databases has largely removed this obstacle and public research challenges have also helped to compare developed approaches on standardized datasets~\cite{Knoll2020a,Muckley2021}. However, they have also highlighted new issues. While it was demonstrated that deep learning models generally outperform handcrafted regularizers in iterative image reconstruction in terms of quantitative metrics like SSIM, PSNR and RMSE, their performance in the regime of over-regularization (when the influence of the prior becomes dominant because of the low-quality of the data) is challenging to assess. The results of classic regularizers like Tikhonov, total variation or $\ell_1$-wavelets in this scenario can be interpreted much easier by end-users. They lead to very characteristic artifact patterns that are easy to spot as being technical artifacts. Deep learning models have the computational capacity to generate realistic-looking images with either missing or artificially hallucinated image features \cite{Hammernik2021} that are inconsistent with the measurement data if they are used at acceleration levels that are too high with respect to the encoding capabilities of the multi-element receive coil. 

A solution is to move from qualitative image assessment towards the assessment of clinical outcomes. In particular, does the diagnostic quality improve for patients when deep learning methods are used instead of handcrafted regularizers? However, conducting such clinical studies is slow and costly, and in many cases an imaging exam cannot even be considered to be a true ground truth, which requires follow-ups with pathology or surgery departments.

Research challenges are also limited in terms of their ability to assess model generalization. The 2020 fastMRI challenge~\cite{Muckley2021} included a track that specifically evaluated generalization with respect to deploying a trained model at a scanner from a different manufacturer. While the winning models performed well in this test, the performance of some approaches was impacted negatively by trivial modifications in the data, for example whether raw data is saved with oversampling in the readout direction or not. In light of the substantial range of imaging parameters that can be changed during an MRI acquisition, it is still an open question if deep learning models should serve a general purpose role, or if specialized models should be tailored to a more narrow range of imaging settings for dedicated exams.

Another open issue of almost all developments is that they are performed with retrospectively accelerated acquisitions, i.e., the accelerated acquisition is obtained by applying a simulated undersampling on the fully sampled dataset. While this is acceptable if true k-space raw data is used and no subtle data crime is performed~\cite{Shimron2021}, not all MR-signal-acquisition effects are captured with retrospective undersampling. In particular, spin-history, gradient and RF-hardware related effects seen in prospectively accelerated acquisitions, i.e. when data is acquired with true undersampling on the scanner, are usually not captured in retrospective acceleration. This can cause issues when moving to prospectively accelerated acquisitions on real MR-scanners. However, it should be pointed out that this is a general issue of all computational imaging methods that are developed retrospectively, and not a unique issue of deep learning techniques.

{While physics-based learning for MRI reconstruction has been successfully established over the past years, there are some potential pitfalls and limitations of these approaches in practice. or instance, when DICOM images are used for experiments instead of raw k-space data, learning-based approaches may lead to overall optimistic results, while real-world unprocessed data performs much worse, resulting in biased state-of-the-art results~\cite{Shimron2021}. In another line of work, the stability of various single-coil networks to small adversarial perturbations at their input was studied ~\cite{Antun2020}, and it was found that networks may exhibit large perturbations at the output. Furthermore, the definition of acceleration factor might also often be misleading. As shown in \cite{Johnson2021}, different sampling patterns yield different results, which opens the question how potential mis-reconstructions can be identified and how the uncertainty of the reconstructions regarding the sampling pattern can be quantified. Finally, we would like to note that the reported undersampling factor and acceleration rates have to be carefully investigated. In the Cartesian setting, undersampling factor and acceleration rates are equivalent and defined by the number of sampled lines divided by the number of total lines in k-space. However, in the non-Cartesian case, the overall effective undersampling depends on the size of oversampling performed along the readout trajectories, without affecting the acceleration factor between the readouts.
}

\subsection{Domain-specific knowledge in post-processing and multi-task imaging}
The medical imaging pipeline consists of many tasks that are mostly viewed separately. The imaging pipeline starts with data acquisition, followed by image reconstruction. The reconstructed image is then further analyzed using post-processing tasks, image segmentation, and quantitative evaluation, and/or methods for diagnosis and treatment planning are applied to facilitate medical decisions.

Thus, there have been efforts to combine several of these tasks into a multi-task imaging framework. Most work on solving multiple tasks jointly has been conducted in the field of motion-corrected image reconstruction, as summarized in Section \ref{sec:nonlinearMotion}. In~\cite{Oksuz2020}, joint motion detection, correction and segmentation was proposed. In contrast to the previously mentioned approaches, the motion was detected directly in k-space and, hence, influence the data consistency layer. Additionally, a bidirectional recurrent CNN (BCRNN) was used to account for spatio-temporal redundancies. The motion-corrected image was obtained by cascading 10 data consistency layers and BCRNNs. Afterwards, a UNet was applied for cardiac segmentation. Evaluated on the UK Biobank data, this work showed that training a joint network for reconstruction and segmentation outperforms sequential training of these networks.

A unified network for joint MRI reconstruction and segmentation was also proposed \cite{Sun2019}. For image reconstruction, an unrolled network with alternating data consistency layers and denoising networks are used. The denoising networks are based on an encoder-decoder structure, where the encoder is shared with the image segmentation network. Hence, common features are extracted using the encoder, while the decoder adapts to the underlying task. Evaluation and simulation of k-space is performed on the MRBrainS segmentation challenge dataset. Their results suggests that high-quality segmentation benefits from this multi-task architecture. A similar conclusion was drawn using FR-Net~\cite{Huang2019}, which consists of two submodules, a reconstruction network based on the fast iterative shrinkage-thresholding algorithm (FISTA), and a UNet for the segmentation of myocardium contours. The networks are trained using a combined reconstruction and segmentation loss, however, finding the right trade-off between these two terms is key for high-quality results. While these approaches both optimize for image reconstruction and another downstream task such as segmentation, it is still an open question if intermediately reconstructed images are needed, or if one could directly obtain, e.g., segmentation in k-space as proposed by~\cite{Schlemper2018segmentation}.

Another interesting line of work incorporates domain knowledge into other post-processing tasks. In deformable image registration, data {similarity} was embedded in an unrolled gradient-descent network~\cite{Qiu2021}, similar to the ideas of  Section \ref{Sec:3d}. This approach only requires the similarity measure to be differentiable.
In medical image segmentation, there has been interest in incorporating prior domain knowledge about the shape, topology or location of the segmented object into the deep learning networks to help with accuracy and improve weak/semi-supervision \cite{eljurdi2021segmentation}. One common way to achieve this is to incorporate the prior domain knowledge into the training loss, similar to the ideas of Section \ref{sec:3a}, including constraints on size, shape and topology of organs incorporated as regularization loss terms \cite{eljurdi2021segmentation}. Another approach is analogous to the ideas of Section \ref{Sec:3d} applied to deformable models for segmentation, where the segmentation starts from a candidate shape that is topologically consistent, and the network learns the appropriate deformation to this shape to best match the label/reference data \cite{clough2019topological}. 

\section{Conclusion}

Physics-driven deep learning techniques are the current state-of-the-art methods for computational MRI. Spanning methods that incorporate physics of MRI acquisitions into loss functions to plug-and-play techniques, and generative models to unrolled networks, a large number of approaches have been proposed to improve the solution of linear and {non-linear} inverse problems that arise in MRI. These methods are starting to make their way into translational and clinical settings, while also potentially altering the downstream tasks in the medical imaging pipeline. Thus, there are numerous opportunities for new technical developments and applications in physics-driven computational MRI from the signal processing community.

\bibliographystyle{IEEEbib}

\bibliography{references}

\end{document}